\documentclass[showpacs,aps,prb,twocolumn,superscriptaddress,preprintnumbers,amsmath,amssymb,floatfix,10pt]{revtex4}
\usepackage{graphicx}
\usepackage[tight]{subfigure}
\usepackage{dcolumn}
\usepackage{float}
\usepackage{bm}
\usepackage{enumerate}
\usepackage[english]{babel}
\usepackage{amsfonts}
\usepackage{amsthm}
\usepackage{amssymb}
\usepackage{calc}
\usepackage{color}
\usepackage{ifthen}

\newcommand{\bfr}{\textbf{r}}
\newcommand{\f}{\frac}
\newcommand{\bfA}{\textbf{A}}
\newcommand{\bfm}{\textbf{m}}
\newcommand{\bfq}{\textbf{q}}
\newcommand{\bfQ}{\textbf{Q}}

\newcommand{\bfJ}{\textbf{J}}

\newcommand{\non}{\nonumber}
\newcommand{\be}{\begin{equation}}
\newcommand{\ee}{\end{equation}}
\newcommand{\ba}{\begin{eqnarray}}
\newcommand{\ea}{\end{eqnarray}}
\newcommand{\bastar}{\begin{eqnarray*}}
\newcommand{\eastar}{\end{eqnarray*}}
\def\comment#1{}

\newcommand{\beg}{\begin{eqnarray}}
\newcommand{\eee}{\end{eqnarray}}

\newcommand{\e}{\mathrm{e}}
\usepackage{inputenc}
\inputencoding{utf8}

\begin{document}
\title{Phase transitions in a three dimensional $U(1) \times U(1)$
  lattice London superconductor: Metallic superfluid and charge-4e
  superconducting states.}
\author{Egil V. Herland}
\affiliation{Department of Physics, Norwegian University of Science and Technology, N-7491 Trondheim, Norway}
\author{Egor Babaev}
\affiliation{Department of Theoretical Physics, The Royal Institute of Technology, 10691 Stockholm, Sweden}
\affiliation{Physics Department, University of Massachusetts, Amherst,
  Massachusetts 01003, USA}
\author{Asle Sudb\o}
\affiliation{Department of Physics, Norwegian University of Science and Technology, N-7491 Trondheim, Norway}

\begin{abstract}
We consider a three-dimensional lattice $U(1) \times U(1)$  and $[U(1)]^N$ superconductors in the London 
limit, with individually conserved condensates. The  $U(1) \times U(1)$ problem, generically, has two types 
of intercomponent interactions of different characters. First, the condensates are interacting 
via a minimal coupling to the same fluctuating gauge field. A second type of coupling is the direct 
dissipationless drag represented by a local intercomponent current-current coupling term in the
free energy functional.
 In this work, we present a study of the 
phase diagram of a $U(1) \times U(1)$ superconductor which includes both of these interactions. 
We study phase transitions and two types of competing paired phases
which occur in this general model: (i) a metallic superfluid phase
(where there is order only in the gauge invariant phase difference of the order
parameters), (ii) a composite superconducting phase where there is order 
in the phase sum of the order parameters which has many properties of 
a single-component superconductor but with a doubled value of electric charge.
We investigate the phase diagram with particular focus on what we call
``preemptive phase transitions.'' These are phase transitions {\it
unique to multicomponent
condensates with competing topological objects}.  A sudden proliferation of
one kind of topological defects may
come about due to a fluctuating background  of topological defects in
other sectors of the theory.
For $U(1) \times U(1)$ theory with unequal bare stiffnesses 
where components are coupled by a non-compact gauge field only, we 
study how this scenario leads to a merger of two $U(1)$ 
transitions into a single  $U(1) \times U(1)$ discontinuous phase transition.
We also report a general form of vortex-vortex bare interaction potential
and possible phase transitions
in an $N$-component London superconductor with individually conserved condensates.

\end{abstract}
\maketitle

\section{Introduction}
Phase diagrams and critical phenomena  in superfluids and superconductors with $U(1)$ symmetry
are well understood theoretically and well investigated numerically. The understanding is largely based
on identifying and describing the behavior of proliferating topological defects.
In two dimensions, a transition from a superfluid to a normal state can be described
as unbinding of vortex-antivortex pairs, which disorders the superfluid phase yielding a 
Berezinskii-Kosterlitz-Thouless transition into a normal state.\cite{kt}
In three dimensions, the topological defects of $U(1)$ theory are vortex loops,
proliferation of which yields a continuous phase transition in the
three dimensional 3Dxy universality class in the case of superfluids (with global $U(1)$
symmetry), or inverted 3Dxy in the case of superconductors (with local $U(1)$ symmetry).\cite{dasgupta,peskin_thomas,fossheimsudbo}
However, it was recently found that in interacting mixtures of $U(1)$ symmetric condensates the situation 
changes principally, yielding much more complex physics, different phase diagrams and transitions.
Many aspects of the phase transitions in systems with several interacting components  are still 
poorly understood and debated.

The main important new aspect arising in an interacting mixture is connected with the fact that, as reviewed 
below, under certain quite generic conditions the vortices with high topological charge (or \textit{bound states of 
vortices}) acquire crucial importance for various aspects in the physics of these systems. This is in contrast to 
single-component systems where only the lowest-topological-charge defects (i.e., only vortices with $2\pi$ phase 
winding) are important. The complexity arising from the relevance of topological defects with high topological charge 
include formation of what is called ``metallic superfluid phases,'' in context of electrically charged  systems, or 
``paired phases,'' in context of electrically neutral systems. In these states no conventional real space pairing takes 
places. However, there is order only in the sum or difference of the phases of the condensate with phases being 
individually disordered.\cite{npb,nature,kuklov,kuklov1,prl05-2,prb05} Moreover, it also results in a complicated and 
still poorly understood nature of the phase transitions from a fully symmetric state to a state with all symmetries 
broken,\cite{motrunich,kuklov,steinar,eskil1} when there is a competition between proliferating low- and 
high- order topological defects.  This is again a phenomenon which has no counterpart in single-component systems.
Various aspects of related effects were also studied in different models with 
a compact gauge field and with SU(2) symmetry. \cite{chernodub}

Recently, it has been found that two kinds of intercomponent interactions lead to the distinct mixture-specific phenomena 
mentioned above. Namely, in a mixture of charged condensates, the intercomponent interaction is represented by the coupling 
between the charged complex scalar matter fields mediated by a fluctuating gauge field.
\cite{npb,nature,kuklov,prl05-1,prl05-2,prb05} On the other hand, in the case of an electrically neutral condensate 
mixture, some related (but at the same time principally different) effects can be produced by a strong dissipationless drag 
(current-current interaction \cite{kuklov1,eskil1,eskil_2_3} which in some physical situations is also called Andreev-Bashkin interaction).\cite{AB} The intercomponent couplings by gauge field and the dissipationless 
drag have so far only been studied separately, while in a generic $U(1) \times U(1)$ system, terms leading to both of these 
effects are allowed by symmetry.  Thus, generically the phase diagram and critical phenomena in a $U(1) \times U(1)$ system 
is a problem with two coupling constants. The interplay between them has, to our knowledge, not been investigated so far.

In this work, we report a quantitative  study of a generic $U(1)\times U(1)$ London superconductor
which has both kinds of intercomponent coupling (gauge field and current-current drag). This includes, in 
particular, the situations where these two different kinds of intercomponent couplings compete with each other. 

This paper is organized as follows. In Sec. \ref{Sec:TheModel}, the general model we consider is introduced, and 
the neutral and charged modes and the vortex representation of the general model, obtained by a duality 
transformation, are identified. Section \ref{Sec:MCDetails} is devoted to the numerical methods we employ in this 
study. The results obtained in the special case with no intercomponent dissipationless drag, is presented in 
Sec. \ref{Sec:MCResults_no_drag}, followed by the results of the general model with competing gauge field and 
Andreev-Bashkin interactions in Sec. \ref{Sec:GeneralResults}. In
Sec. \ref{Sec:N-comp_case}, we discuss the general $N$-component case,
before we arrive to the conclusions in Sec. \ref{Sec:Conclusion}. We also present analytical details presenting the duality transform for a general $N$-component 
model in Appendix \ref{App:N_comp_dualization} and a derivation of the expression for the gauge field correlator in 
Appendix \ref{App:gauge_field_corr}.  

\section{Model \label{Sec:TheModel}}
We study a generic two-component London superconductor. In the London limit, one neglects the 
fluctuations of the density fields $|\psi_{1,2}|$ of the complex scalar functions  $\psi_j=|\psi_j|\e^{i\theta_j}$  describing two superconducting components (i.e., setting 
$|\psi_{1,2}| \approx const$). Fluctuations of the phases $\theta_j$, and the gauge field ${\bf A}$
are allowed. The compact support of the phase-variables $\theta_i \in [0, 2 \pi \rangle$ implies 
that phase-fluctuations lead to vortex-excitations, capable of destroying 
superconductivity/superfluidity, in this system. The London limit is an adequate approximation 
for many properties of strongly type-II superconductors, and in fact transcends the validity of 
the Ginzburg-Landau theory. The free-energy density of this system can be written as
\begin{equation}
\begin{split}
 F &= \sum_{j=1,2} \frac{\rho_j}{2}(\nabla
   \theta_j-e_j \bfA)^2 + \frac{(\nabla
     \times \bfA)^2}{2}\\  &\quad - \frac{\rho_d}{2}(\nabla
   \theta_1 - e_1 \bfA - \nabla \theta_2 + e_2 \bfA)^2,
\label{Eq:london}
\end{split}
\end{equation}
where $\rho_j$ physically represent the bare phase stiffnesses of the problem. 
In addition to the intercomponent 
coupling between the two charged condensates via a fluctuating gauge field ${\bf A}$, we include a 
direct intercomponent dissipationless current-current interaction with strength $\rho_d$, which has 
the form\cite{AB}
\be 
F_{drag}=\rho_d(\nabla  \theta_1 - e_1 \bfA) \cdot(\nabla \theta_2 - e_2 \bfA). 
\ee
It is a part of the last term in \eqref{Eq:london}. The particle currents of both species then 
depend on the common vector potential and superfluid velocities of
both condensates (i.e., particles belonging to one condensate can be carried by superfluid velocity of the other),
\begin{align}
{\bf j}_1 = (\rho_1 - \rho_d) (\nabla \theta_1 - e_1\bfA) +
\rho_d(\nabla \theta_2 - e_2\bfA),\\
{\bf j}_2 = (\rho_2 - \rho_d) (\nabla \theta_2 - e_2\bfA) + \rho_d(\nabla \theta_1 - e_1\bfA). 
\end{align}

For generality, we allow for unequal charges $e_j$ in the two components of the system, examples of the 
systems with oppositely charged condensates are given below. Note that the drag term implies that there is a stability criterion 
that must be applied to the system. If $\rho_d$ exceeds a critical limit, to be determined below, the spectrum of the system will 
be unbounded from below and hence the theory will be ill defined. The bare stiffness coefficients $\rho_j$ must be 
positive, $\rho_j \geq 0$, on simple physical grounds.

The physical model in Eq. \eqref{Eq:london} is discussed in the context of the projected quantum ordered states of 
hydrogen or its
isotopes at high compression \cite{nature,prb05,prl05-1,prl05-2,nphys,prl02} where the different fields correspond
to condensates formed by electrons, protons or deuterons.
A similar model appears in some models of neutron stars interior where the two fields represent protonic and 
$\Sigma^-$ hyperon Cooper pairs.\cite{Jones_prl09ns} 
Moreover, the model with equal phase stiffnesses 
$\rho_1 = \rho_2$ and charges $e_1 = e_2$, appears as an effective model in the theories of easy-plane quantum 
antiferromagnets.\cite{DQC1,DQC2} Related models were also studied in various contexts in low dimensions.\cite{npb,2d}

The model has topological excitations which are vortices
with $2\pi n_j, \ n_j= \pm 1, \pm 2, ...$ phase winding in the phase of component $j$. We denote vortices
by the pair of integers $(n_1,n_2)$ characterizing phase windings of the vortex in
question. Thus, vortices with phase winding in only one component are
denoted $(1,0)$ or $(0,1)$. The model also possesses composite vortices where both integers associated with the phase windings (around or nearly 
around the same core) in the two species are nonzero. In this paper, we will only consider the composite vortices
$(1,1)$ and $(1,-1)$ which have co-directed and counter-directed phase windings in the two components, respectively. However, composite 
vortices with higher topological charges, such as $(1,n_2)$ or $(n_1,1)$, may be relevant under certain 
conditions.\cite{eskil1, Kaurov_05}

\subsection{Charged and neutral modes}
\label{Subsec:Charged_neutral}
By separation of variables,\cite{prl02,Babaev_PRB_02,prb05} we may rewrite the model in
Eq. \eqref{Eq:london} in a form where the composite charged and neutral modes are
explicitly identified,
\begin{equation}
\begin{split} 
\!\!\!\!F =
\frac{1}{2}\Bigg[&\frac{\rho_1\rho_2-\rho_d(\rho_1+\rho_2)}{m_{0}^{2}}(e_2\nabla
\theta_1-e_1\nabla \theta_2)^2\\ 
&+ \frac{1}{m_{0}^{2}}(e_jR_{j1}\nabla
  \theta_1 + e_jR_{j2}\nabla \theta_2 - m_{0}^{2}{\bf A})^2\\
&+ (\nabla \times {\bf A})^2 \Bigg],
\end{split}
\label{Eq:Charged_and_neutral_modes}
\end{equation} 
where the coefficients are given by 
\begin{equation}
R = \begin{pmatrix} \rho_1-\rho_d & \rho_d \\
\rho_d & \rho_2-\rho_d
\end{pmatrix}
\label{Eq:R_matrix}
\end{equation}
and
\begin{equation}
m_{0}^{2} = (\rho_1-\rho_d)e_1^2 + (\rho_2-\rho_d)e_2^2 +
2\rho_de_1e_2.
\label{Eq:M}
\end{equation}
Throughout the paper, there is an implicit sum over repeated component indices. The coefficient $m_{0}$ should not be confused with the mass
of the components. These are included in $\rho_j$, whereas $m_{0}$
determines the inverse bare screening length of the screened
interactions in the system, details will be given in Sec. \ref{Subsec:Dual_model}.
The first term of Eq. \eqref{Eq:Charged_and_neutral_modes} is identified
as the neutral mode that does not couple to the vector potential. The second
term is the charged mode, characterized by its coupling to the vector potential.

From Eq. \eqref{Eq:Charged_and_neutral_modes}, it is seen that for stability of the system
(in the sense that the free energy  functional should be bounded from below)
the coefficient of the first term should be positive. It is readily shown that the criterion 
for this is that
\begin{eqnarray}
\rho_d < \frac{\rho_1 ~ \rho_2}{\rho_1 + \rho_2}.
\label{Stability}
\end{eqnarray}
Note that this criterion is identical to the one derived in Ref. \onlinecite{eskil1},
and does not depend on charge.  Actually, there are no restrictions on
the value of the electric charge $e$, to obtain a well-defined
theory.   
 
Note that in Eq. \eqref{Eq:london}, the phases 
of the two components do not represent gauge invariant quantities. However, when the model is 
rewritten on the form in Eq. \eqref{Eq:Charged_and_neutral_modes}, observe that the neutral 
mode identifies a linear combination of the phase gradients that is a gauge invariant
quantity decoupled from the vector potential ${\bf A}$, 
$\nabla(e_2\theta_1-e_1\theta_2)$.  Thus, the $U(1)\times U(1)$ symmetry of the model 
may be interpreted  as possessing a ``composite'' 
electrically neutral (or ``global'')  $U(1)$ symmetry associated 
with the phase combination of the neutral mode, and a composite $U(1)$ gauge 
symmetry which is coupled to vector potential ${\bf A}$ and thus is
associated with the charged mode.     
Importantly, the identification of the charged and a 
neutral mode does not imply that the modes are decoupled, because both modes
depend on phases $\theta_i$ which are constrained to have $2\pi \times {\rm integer}$ 
phase windings.

\subsection{Case $\rho_d=0$, $e_1=e_2=e$\label{Subsec:Easy_plane_case}}
We briefly review the physics of a two-component $U(1) \times U(1)$ superconductor 
with individually conserved condensates, coupled only by the gauge field,
i.e., in the absence of Andreev-Bashkin (i.e., mixed gradient) terms.
In the London limit the free energy may be read off from Eq. \eqref{Eq:Charged_and_neutral_modes},
\begin{equation}
\begin{split} 
F=&\frac{1}{2} \f{\rho_1\rho_2}{\rho_1+\rho_2}
[\nabla (\theta_1-\theta_2)]^2\\ &+ \frac{1}{2} 
\f{ \left[\rho_1\nabla \theta_1 +\rho_2\nabla \theta_2- e (\rho_1+\rho_2) \bfA \right]^2 }{\rho_1+\rho_2}\\
&+ \f{1}{2} (\nabla \times \bfA)^2.
\end{split}
\label{gl2}
\end{equation} 

The important new physics arising in the model, Eq. \eqref{gl2}, compared to single-component GL model is that the 
lowest-order topological defects with a $2\pi$ phase winding only in one phase $\theta_i$ have a logarithmically 
diverging energy per unit length due to a neutral supercurrent, while vortices where both phases have $2\pi$ winding have finite energy per unit 
length.\cite{prl02,babaev1} Under certain conditions vortices where both phases wind, i.e., (1,1), can proliferate 
without triggering a proliferation of the simplest vortices $(1,0)$ and (0,1).

Consider now a composite (1,1) vortex. Such an excitation, if vortices in two components share the same core, has 
nontrivial contribution to the following terms in the free energy, Eq. (\ref{gl2}):
\begin{equation}
\begin{split}
F_{(1,1)}^{\textnormal{eff}}=&\frac{1}{2} \f{ \left[ \rho_1\nabla
    \theta_1 + \rho_2\nabla \theta_2 - e (\rho_1+\rho_2) \bfA  \right]^2
}{\rho_1+\rho_2}\\
&+ \f{1}{2} (\nabla \times \bfA)^2.
\end{split}
\label{gl3}
\end{equation}
If the (1,1) vortex has phase windings around a common core, it can be mapped onto a vortex in a single component 
superconductor. Then, by increasing electric charge one can make the energy cost of a vortex per unit length in a 
lattice London superconductor arbitrarily 
small (because the vortex energy depends logarithmically on the penetration depth which is in turn a function of electric 
charge). Thus, in a lattice London superconductor the critical temperature of proliferation of the vortices can be arbitrary 
small if the value of the electric charge is sufficiently
large. Therefore, in the two-component model, Eq. (\ref{gl2}), one may, by 
increasing the value of electric charge, proliferate (1,1) vortices without proliferating individual vortices (1,0) or 
(0,1). The latter two produce a phase gradient in the gauge invariant phase difference $\theta_1-\theta_2$. This features 
a stiffness which is not renormalized by the proliferation of the (1,1) vortices.

Since the (1,1) vortices do not 
have a topological charge in the phase difference, they cannot
disorder the first term in Eq. (\ref{gl2}),  but they 
disorder the charged sector represented by the second term. The resulting state therefore features long-range
ordering in the phase difference and can be characterized by
$\langle\e^{i(\theta_1-\theta_2)}\rangle \ne 0$, while
$\langle\e^{i\theta_1}\rangle = 0$, $\langle\e^{i\theta_2}\rangle = 0$, and there is 
no Meissner effect. The free energy for the resulting phase is given
by the following term (i.e. it has only broken global $U(1)$ symmetry)
while the stiffness of the charged $U(1)$ mode is renormalized 
to zero by proliferated composite vortices,
\begin{equation}
F_{(1,-1)}^{\textnormal{eff}}=\frac{1}{2} \f{\rho_1\rho_2}{\rho_1+\rho_2}
[\nabla (\theta_1-\theta_2)]^2.
\label{glms}
\end{equation}
The proliferation of composite defects resulting into this state was shown to arise in two-dimensional (2D)
systems at any finite temperatures.\cite{npb} In three dimensions, this phase can be induced by a magnetic field 
via melting of a composite vortex lattice.\cite{nature,prl05-2} An analogous phase was also found in a three
dimensional lattice superconductor arising without applied external
field from fluctuations if the value of the 
electric charge is very large.\cite{kuklov} Since there is no Meissner effect in the resulting phase, but at 
the same time there is a broken neutral $U(1)$ symmetry, the term metallic superfluid (MSF) was coined for 
it.\cite{nature} Also related phases are sometimes called ``paired phases.''\cite{kuklov} The latter term is 
motivated by the fact that in such situations the (quasi-) long-range order is retained only in some linear 
combination of phases while individual phases are disordered. Importantly it should not be confused with the  
conventional ``real-space'' pairing of bosons.

\subsubsection{Case $\rho_1=\rho_2$}
Consider the case where $\rho_1=\rho_2$. At high values of the electric charge $e$, the model was shown to feature a MSF 
phase without applied field.\cite{kuklov} This implies that at large $e$ the system undergoes two phase transitions 
when the temperature is increased. The first transition is 
from a state with broken $U(1)\times U(1)$ symmetry into the MSF  with broken $U(1)$ symmetry,
driven by a proliferation of composite (1,1) vortices. The second transition 
is one where the remaining broken $U(1)$ symmetry is restored by proliferation
of individual vortices, resulting in a normal state. At low values of $e$,  
one cannot separate characteristic temperatures of the proliferation of composite and individual vortices and 
thus, the model should have only one phase transition from broken $U(1)\times U(1)$ to a normal state. In the case 
$\rho_1=\rho_2$, the latter phase transition was conjectured to be a continuous phase transition in a novel 
universality class in the work of Ref. \onlinecite{motrunich}. However, subsequent works show
that the phase transition is first order,\cite{steinar,kuklov} see
also Ref. \onlinecite{chernodub}. 
Moreover, the analysis performed in Ref. \onlinecite{kuklov} indicates that the  $U(1)\times U(1)$ to a normal state 
transition is first order for any values of electric charge in the $\rho_1=\rho_2$ model. Note that the standard 
theories of vortex loop proliferation 
yield a second order transition.\cite{dasgupta,fossheimsudbo} An analysis of a simpler two-component model (with no 
gauge field coupling, but with direct current-current coupling) which,
like the model, Eq. (\ref{gl2}), also features
low-energy composite vortices, provides some evidence that the first order transition takes place whenever a 
restoration of the $U(1)\times U(1)$ broken symmetry is driven by proliferation of  {\it competing tangles of different 
kinds of
vortices},\cite{eskil1} e.g., tangles of (1,0), (0,1) vortices  and  a
tangle of (1,1) vortices. The term ``preemptive vortex-loop
proliferation transition'' was coined for this scenario.\cite{eskil1}
Note that in a charged $U(1) \times U(1)$ theory for arbitrary values of
electric charge one cannot rule out  
in a simple way  that composite vortices participate in a competition
with the individual vortices in the symmetry-restoration transition
since composite vortices have finite energy per unit length.

\subsection{Dual model \label{Subsec:Dual_model}}
We will now perform a duality transformation that reduces the model in
Eq. \eqref{Eq:london} to a theory of interacting 
vortex loops of two species. These are the topological objects which drive the phase transition between the normal 
state and a state with broken symmetries in the systems we consider. 
When the phases and gauge field are fluctuating
the statistical sum of the London two-component 
superconductor with intercomponent drag can be represented as follows:
\begin{gather}
Z = \int \mathcal{D}\theta_1 \int \mathcal{D}\theta_2 \int \mathcal{D}
\bfA \quad \e^{-S},\non
\\ 
\begin{split}
S &= \frac{\beta}{2}\int d^3r \Big\{[\nabla \times \bfA(\bfr)]^2\\
&\quad + \left[\nabla
  \theta_j(\bfr)-e_j \bfA(\bfr)\right]R_{jk}\left[\nabla
  \theta_k(\bfr)-e_k \bfA(\bfr)\right]\Big\},
\label{Eq:Dualization1_S}
\end{split}
\end{gather}
where $\beta$ is the inverse temperature.

We now choose the gauge $\nabla \cdot \bfA(\bfr) = 0$ and Fourier transform the action. The action is then written as
\begin{equation} 
\begin{split}
S = \frac{\beta}{2} &\int d^3q \Bigg[\tilde{\bfA}(\bfq)(q^2+m_{0}^{2})\tilde{\bfA}(-\bfq)\\
&+ \textbf{U}_j(\bfq)\left(R_{jk}-\frac{e_le_mR_{lj}R_{mk}}{q^2+m_{0}^{2}}\right)\textbf{U}_k(-\bfq)\Bigg], 
\label{Eq:Dualization2_S}
\end{split}
\end{equation}
where the Fourier transform of $\nabla \theta_j(\bfr)$ is denoted by $\textbf{U}_j(\bfq)$. Moreover we have completed 
the squares of the gauge field with $\tilde{\bfA}(\bfq) = \bfA(\bfq)-e_{j}R_{jk}\textbf{U}_k(\bfq)/(q^2+m_{0}^{2})$ as the shifted 
gauge field. By integration of the shifted gauge field, the model is
written as
\begin{gather}
Z = \int \mathcal{D}\theta_1 \int \mathcal{D}\theta_2\quad \e^{-S},\non\\
S = \frac{\beta}{2} \int d^3q \;\textbf{U}_j(\bfq)\left(R_{jk}-\frac{e_le_mR_{lj}R_{mk}}{q^2+m_{0}^{2}}\right)\textbf{U}_k(-\bfq), 
\label{Eq:Dualization3_S}
\end{gather}
with the phases as the only remaining fluctuating quantities. The phase gradient can be decomposed into 
a longitudinal and a transverse part, $\textbf{U}_j(\bfq) = [\textbf{U}_j(\bfq)]_L + [\textbf{U}_j(\bfq)]_T$,
where the longitudinal component corresponds to regular smooth phase-fluctuations with zero curl, i.e., "spin 
waves.'' Hence, the longitudinal part is curl free, $\bfq \times [\textbf{U}_j(\bfq)]_L = \textbf{0}$ and the 
transverse part is divergence free, $\bfq \cdot [\textbf{U}_j(\bfq)]_T = 0$ and thus it is associated with
quantized vortices. One can introduce the field $\bfm_j(\bfq)$ which
is the Fourier transform of the integer-valued vortex field 
for component $j$,
\begin{equation}
i\bfq \times [\textbf{U}_j(\bfq)]_T = 2\pi \bfm_j(\bfq), \qquad j = 1\textnormal{, }2. 
\label{Eq:Curl_cond}
\end{equation}
Note that this relation yields the constraint $\bfq \cdot \bfm_j(\bfq) = 0$,
i.e., the thermal vortex excitations in the theory are closed loops as required
by the single-valuedness of the order parameter in an infinite system. In the following, we will disregard 
the longitudinal phase fluctuations since the physics at the  critical points in this system is governed by the vortex 
excitations and not the spin waves. The latter are known to be innocuous and incapable of destroying
long-range order in three dimensional systems. By Eq. \eqref{Eq:Curl_cond}, the transverse phase 
gradient is explicitly written as
\begin{equation}
[\textbf{U}_j(\bfq)]_T = 2\pi i \frac{\bfq \times \bfm_j(\bfq)}{q^2},
\qquad j = 1\textnormal{, }2, 
\label{Eq:Transversal_part_fourier_rep}
\end{equation} 
and thus, we  finally express the statistical sum via vortex fields, 
\begin{gather}
Z = \sum_{\bfm_1}\sum_{\bfm_2}\quad \e^{-S},\non\\
S = 2\beta \pi^2 \int d^3q \; \bfm_j(\bfq)V_{jk}(q^2)\bfm_k(-\bfq).
\label{Eq:Dualization4_S}
\end{gather}
Here, the summation over the vortex fields $\bfm_j(\bfq)$ is
constrained by $\bfq \cdot \bfm_j(\bfq) = 0$, such that the
integer-valued real-space vortex fields $\bfm_j(\bfr)$
form closed loops only.
The vortex-vortex interactions are given by
\begin{widetext}
\begin{equation}
V_{jk}(q^2) = \frac{1}{q^2}  \left( R_{jk} - \frac{e_le_mR_{lj}R_{mk}}{q^2+m_{0}^{2}}
\right) \Leftrightarrow V(q^2) = 
\begin{pmatrix}
\dfrac{\rho_1-\rho_d-\frac{(e_{j}R_{j1})^2}{m_{0}^{2}}}{q^2} +
\dfrac{\frac{(e_{j}R_{j1})^2}{m_{0}^{2}}}{q^2+m_{0}^{2}} &
\dfrac{\rho_d-\frac{e_{j}e_{k}R_{j1}R_{k2}}{m_{0}^{2}}}{q^2} +
\dfrac{\frac{e_{j}e_{k}R_{j1}R_{k2}}{m_{0}^{2}}}{q^2+m_{0}^{2}}\\
\dfrac{\rho_d-\frac{e_{j}e_{k}R_{j1}R_{k2}}{m_{0}^{2}}}{q^2} +
\dfrac{\frac{e_{j}e_{k}R_{j1}R_{k2}}{m_{0}^{2}}}{q^2+m_{0}^{2}} & 
\dfrac{\rho_2-\rho_d-\frac{(e_{j}R_{j2})^{2}}{m_{0}^{2}}}{q^2} + \dfrac{\frac{(e_{j}R_{j2})^{2}}{m_{0}^{2}}}{q^2+m_{0}^{2}} 
\end{pmatrix}.
\label{Eq:Dual_V}
\end{equation}
\end{widetext}
Here, we have used the identity
\begin{equation}
[\textbf{U}_j(\bfq)]_T\cdot [\textbf{U}_k(-\bfq)]_T = \frac{(2\pi)^2}{q^2}\bfm_j(\bfq)\cdot\bfm_k(-\bfq),
\end{equation}
found by Eq. \eqref{Eq:Transversal_part_fourier_rep}. We may now
interpret $m_{0}$, given by Eq. \eqref{Eq:M}, 
as the inverse bare screening length that sets the scale of the Yukawa interactions in the system. 

We remind the reader briefly of what is known for the one-component
case, i.e., $\rho_2=0$, $\rho_d=0$, $e_2=0$, $\rho_1=\rho \neq 0$, $e_1=e \neq
0$ in Eq. \eqref{Eq:Dual_V}. 
Then, we have $V_{11}= [\rho - e^2 \rho^2/(q^2+m_0^2)]/q^2$, with
$m_0^2= \rho e^2$. Thus, $V_{11}= \rho /(q^2+m_0^2)$ 
is a screened interaction between the vortices, mediated by the fluctuating gauge field.
This is drastically different from the multi-component case,
where one fluctuating gauge field is incapable of fully screening
interactions between vortex excitations in all condensate
fields.\cite{npb, prb05, prl05-2} 

The interactions between vortex elements in the 
system are generally seen to include two parts: A long-range Coulomb interaction with no
intrinsic length scale that decays as $1/r$ and a short-range Yukawa interaction with an exponential 
decay. Note that in the index representation of Eq. \eqref{Eq:Dual_V}, the
first term, $R_{jk}/q^2$ will dominate the second term,
$e_le_mR_{lj}R_{mk}/[q^2(q^2+m_{0}^{2})] \sim q^{-4}$, at short distances when
$q^2$ is large, because the Yukawa and Coulomb part of the second term
will cancel each other. Effectively, at short distances, the vortices will interact as if the gauge field does not fluctuate. On the other hand, at large distances, when $q^2$
is small, the Coulomb part of the second term will dominate its Yukawa
counterpart and the second term will be of the same order as the first
term $\sim q^{-2}$. Thus, the $1/r$ contributions from the gauge field mediated interactions between vortices
sets in when intervortex separation becomes larger than the characteristic
distance $m_{0}^{-1}$. Also note that by decreasing the gauge field coupling constant $e$, $m_{0}^{-1}$ grows 
and so does the distance where the effects of the gauge field are
negligible. In particular, 
when $\rho_d=0$ in Eq. \eqref{Eq:Dual_V} (this corresponds to the work in
Refs. \onlinecite{npb,prb05}), we have the  case that the 
interactions between elementary vortices of different species tend to cancel out at short intervortex
separations, whereas there will be interactions at large intervortex
separations that are mediated by the gauge field.

In the general model with the mixed-gradient terms considered here (i.e., with $\rho_d \ne 0$), there 
is in addition unscreened
$1/r$  interaction between vortices belonging to different condensates
which is mediated by the direct Andreev-Bashkin drag. Thus, contrary to
the $\rho_d=0$ case, there will
be unscreened Coulomb interactions at all length scales. 

Observe that in the limit, $e_1 = e_2 = 0$,  Eq. \eqref{Eq:Dual_V} eliminates 
Yukawa-type interaction potential and resulting to only long-range 
interactions $V(q^2) = R/q^2$ like in a two-component superfluid with Andreev-Bashkin effect, see Ref. \onlinecite{eskil1}. 
Observe also that  
in contrast to the neutral model in Ref. \onlinecite{eskil1}, in the above case
when $e_{1,2} \ne 0$ one always has a bound state of 
vortices which has finite energy per unit length, as discussed in Sec. \ref{Subsec:Easy_plane_case}.

Thus, the vortex-vortex interaction matrix shows that adding the mixed gradient Andreev-Bashkin-type
drag term to a superconductor, where components interact only via a fluctuating gauge field, 
might significantly alter the physics of fluctuations as a consequence of 
a substantial change in the interactions between topological 
excitations.

Finally, in the spirit of Sec. \ref{Subsec:Charged_neutral}, we may rewrite the action in
Eq. \eqref{Eq:Dualization4_S} in a form where the charged and the neutral
modes are explicitly identified,
\begin{widetext}
\begin{equation}
\begin{split}
S = 2\beta \pi^2 \int
d^3q\Bigg\{&\frac{\rho_1\rho_2-\rho_d(\rho_1+\rho_2)}{m_{0}^{2}}\left[e_2\bfm_1(\bfq) -
e_1\bfm_2(\bfq)\right]\frac{1}{q^2}\left[e_2\bfm_1(-\bfq) -
e_1\bfm_2(-\bfq)\right]\\ &+ \frac{1}{m_{0}^{2}}\left[e_{j}R_{j1}\bfm_1(\bfq) + e_{j}R_{j2}\bfm_2(\bfq)\right]\frac{1}{q^2+m_{0}^{2}}\left[e_{k}R_{k1}\bfm_1(-\bfq) + e_{k}R_{k2}\bfm_2(-\bfq)\right]\Bigg\}. 
\end{split}
\label{Eq:neutral_charged_vortex_modes}
\end{equation}
\end{widetext}
Note that the vortex fields in the neutral sector interacts by an unscreened Coulomb interaction only, while
the vortex fields in the charged sector interacts by a screened
Coulomb (Yukawa) interaction. From this it follows that the
corresponding propagators are given by $\langle [e_2\bfm_1(\bfq) - e_1\bfm_2(\bfq)]\cdot [e_2\bfm_1(-\bfq) - e_1\bfm_2(-\bfq)]\rangle \sim q^2$
and $\langle [e_{j}R_{j1}\bfm_1(\bfq) + e_{j}R_{j2}\bfm_2(\bfq)]\cdot[e_{j}R_{j1}\bfm_1(-\bfq) + e_{j}R_{j2}\bfm_2(-\bfq)]\rangle \sim
q^2+\tilde{m}_0^2$. Here, $\tilde{m}_0$ is the {\it effective} dynamically generated 
gauge mass that is nonzero in the low-temperature phase and vanishes at the charged
critical point. Moreover, there is also a neutral critical
point associated with ordering the neutral sector of
Eq. \eqref{Eq:neutral_charged_vortex_modes}
 with a corresponding non-analytic variation in the temperature
dependence of the coefficient of the $q^2$ term. 

Note that for any value of $\rho_d$, the interactions of the vortex
fields in the neutral sector are independent of any variation in the
charges $e_1$ and $e_2$ 
provided that the ratio  
$e_2/e_1$ is kept fixed, as readily seen by inspection of
Eq. \eqref{Eq:neutral_charged_vortex_modes}. On the other hand, the
interactions in the charged sector depends 
on the value of the charge in the Yukawa factor $1/(q^2+m_{0}^{2})$.
 
Given the very different form of intervortex interactions produced by the gauge-field coupling 
and by the Andreev-Bashkin drag, the interesting case when these interactions compete with each 
other cannot be mapped onto the previously studied regimes of systems interacting only by 
gauge field or only by intercomponent drag. Investigating the physics arising from this competition 
 is the main objective in this paper.

\section{Details of the Monte Carlo simulations \label{Sec:MCDetails}}
 
Large-scale Monte Carlo (MC) simulations were performed in order to
explore the phases and phase transitions of the model, Eq. \eqref{Eq:london}. We discretize space into a three-dimensional cubic lattice of size $L \times L \times L$ with 
lattice spacing $a = 1$. The phases are defined on the vertices of the lattice, $\theta_j(\bfr) \rightarrow \theta_{\bfr, j}$ 
and the phase gradient is a finite difference of the phase at two neighboring lattice points, $\partial_{\mu}\theta_j(\bfr)
\rightarrow \Delta_{\mu} \theta_{\bfr, j} =
\theta_{\bfr+\bm{\hat{\mu}}, j} - \theta_{\bfr, j}$. The gauge field is associated with the links between the lattice points, 
$A_{\mu}(\bfr) \rightarrow A_{\bfr, \mu}$. Moreover, the curl of the gauge field yields a plaquette sum 
$(\nabla \times \bfA (\bfr))_{\mu} \rightarrow \sum_{\nu \eta}\varepsilon_{\mu \nu
    \eta} \Delta_{\nu} A_{\bfr, \eta}$. Here, $\varepsilon_{\mu \nu
  \eta}$ is the Levi-Civita symbol. The compact phases $\theta_{\bfr,
  j}$  have to be $2\pi$ periodic. This is accommodated by the Villain approximation 
of the effective  Hamiltonian,\cite{Villain_75} which also yields a faithful lattice representation of the 
direct current-current interaction (i.e., drag) term.\cite{eskil1} Our effective lattice model thus reads
\begin{gather}
Z = \int_{0}^{2\pi} \mathcal{D}\theta_1 \int_{0}^{2\pi} \mathcal{D}\theta_2 \int_{-\infty}^{\infty} \mathcal{D}
\bfA \quad \e^{-\beta
  H[\theta_1, \theta_2, \bfA; \beta]},\non
\\ 
H[\theta_1, \theta_2, \bfA; \beta] = \sum_{\bfr,
  \mu}-\beta^{-1} \ln \left\{ \sum_{n_{\bfr,
  \mu, 1}} \sum_{n_{\bfr, \mu, 2}} \e^{-S} \right\},
\label{Eq:Villain_H}
\end{gather}
where the local Villain action is 
\begin{equation}
\begin{split}
S = \frac{\beta}{2} &\Bigg[\rho_1 u_{\bfr, \mu, 1}^2 + \rho_2 u_{\bfr, \mu, 2}^2
  - \rho_d(u_{\bfr, \mu, 1}-u_{\bfr, \mu, 2})^2\\ 
&\quad+ \left(\sum_{\nu \eta}\varepsilon_{\mu \nu
    \eta} \Delta_{\nu} A_{\bfr, \eta}\right)^{2} \Bigg]. 
\label{Eq:Villain_S}
\end{split}
\end{equation}
Here, $u_{\bfr, \mu, j} = \Delta_{\mu}\theta_{\bfr, j} - e_j A_{\bfr,
  \mu} - 2\pi n_{\bfr,\mu, j}$ is a one-component Villain argument. The sum over the integer-valued fields,
$n_{\bfr,\mu, j}$, is from $-\infty$ to $\infty$ ensures $2\pi$
periodicity of the Hamiltonian with respect to the
  gauge-invariant phase difference.

All Monte Carlo simulations start with an initialization of the system, either disordered, when all phases and 
gauge fields are chosen at random, or ordered, when phases and gauge fields are chosen constant 
throughout the system. Subsequently, a sufficiently large number of sweeps is performed in order to thermalize 
the system. As a valuable check on the simulations, the calculated quantities should be invariant with respect 
to the initialization procedure. A Monte Carlo sweep includes local updating of all five fluctuating field 
variables (compact phases $\theta_{\bfr, j} \in [0,2\pi \rangle$ and the
non-compact gauge field $A_{\bfr, \mu}$)
at all lattice sites in the system, according to the Metropolis-Hastings algorithm.\cite{Metropolis_53_Hastings_70} 
There is no gauge fixing involved, as summation over gauge equivalent 
configurations will cancel out when calculating thermal averages of gauge invariant quantities. Moreover, 
periodic boundary conditions are applied in all simulations. 

In most cases, we also apply the so-called parallel tempering algorithm,\cite{Hukushima_96_Earl_05} allowing a global 
swap of configurations between neighboring couplings, after the local updating is finished. The explicit temperature 
dependence in the Hamiltonian of the Villain model\cite{Janke_93_Kleinert_89} must be considered when calculating 
the probability of exchanging configurations between two coupling values $\beta$, $\beta'$, which is   
\begin{equation}
\label{Eq:PT_move}
W_{PT} = \left\{ \begin{array}{lr}
1, & \textnormal{if }\Delta < 0,\\
\e^{-\Delta}, & \textnormal{if }\Delta \ge 0,\\
\end{array} \right.
\end{equation}
where $\Delta = \beta'(H[X;\beta']-H[X'; \beta']) - \beta(H[X; \beta]-H[X'; \beta])$, and $X$, $X'$ are 
the configurations at $\beta$, $\beta'$ initially. To increase the performance of the
parallel tempering algorithm, the set of coupling values was selected according to the initialization 
procedure in Ref. \onlinecite{Hukushima_99}, to yield approximately the same acceptance rate for the 
parallel tempering move throughout the entire range of coupling values in the simulation. By introducing 
the parallel tempering algorithm, the quality of the statistical output was substantially improved by 
reducing the autocorrelation time at critical points by 1-2 orders of magnitude compared with conventional Monte Carlo 
simulations with local updates only. Even in regions of the phase diagram where coupling intervals were 
too large for configurations to access all coupling values within a reasonable amount of MC sweeps, which is 
required to take full advantage of the parallel tempering method,\cite{Hukushima_96_Earl_05} an improvement 
of the statistical output was achieved.

\subsection{Specific heat}
We measure the specific heat per site $C_v$ by the energy fluctuations, 
\begin{equation}
\frac{C_v L^3}{\beta^2} = \left \langle (H - \langle H\rangle)^2 \right \rangle, 
\label{Eq:specific_heat}
\end{equation}
where the brackets denote thermal average with respect to the
partition function in Eq. \eqref{Eq:Villain_H}. In fact, this expression is not quite right for the
Villain model because of the explicit temperature dependence in the
Hamiltonian.\cite{Janke_93_Kleinert_89} Generally, the specific heat
is given by $L^3C_v = -\beta^2 \partial U/(\partial \beta)$, where the
internal energy is given by $U = - \partial \ln Z/(\partial
\beta)$.\cite{Nguyen_98_2} Thus, the specific heat is written as
\begin{equation}
\frac{C_v L^3}{\beta^2} = \left\langle \left(\frac{\partial (\beta
      H)}{\partial \beta} - \left\langle \frac{\partial (\beta
      H)}{\partial \beta}\right\rangle \right)^2 -
\frac{\partial^2 (\beta
      H)}{\partial \beta^2} \right\rangle. 
\label{Eq:real_specific_heat}
\end{equation}
We expect no extra singular behavior due to the temperature
dependence in the Villain Hamiltonian, so the singular behavior in
Eq. \eqref{Eq:real_specific_heat} should also be captured in the
energy fluctuations of Eq. \eqref{Eq:specific_heat}. Thus, we expect
Eq. \eqref{Eq:specific_heat} to reproduce the correct critical behavior of the
heat capacity, as was the case in Ref. \onlinecite{Nguyen_98_1}. In
practice, both equations were used, and the results were identical
with respect to critical behavior. In the analysis of the Monte Carlo simulations, the critical temperature of the phase
transitions was determined by locating the anomaly of the heat
capacity, and the same critical temperature was found with both equations. 

\subsection{Helicity modulus}
The helicity modulus is a global measure of phase coherence in a 
superfluid (i.e. decoupled from gauge field) order parameter. It measures the energy cost
associated with an infinitesimal twist $\bm{\delta}$ in the phase of an order
parameter across the system. In order to obtain the correct energy cost with respect to
composite phase combinations such as e.g. phase difference, we must 
perform a general twist in a linear combination of the order parameter phases, 
\begin{equation}
\theta_{\bfr, j} \rightarrow \theta_{\bfr,
  j}' = \theta_{\bfr, j}
- a_j \bm{\delta}\cdot \bfr,
\label{Eq:twist}
\end{equation}
where $a_j$ now is a real number associated with the phase twist in
component $j$. By selecting $a_1$, $a_2$, 
we may measure the phase coherence of any linear combination, $a_1 \theta_1 + a_2 \theta_2$, 
in order parameter space. That is, if we want to measure the helicity modulus of the neutral
mode associated with the phase difference we must impose a twist 
in the phase difference, i.e., $a_1 = 1$, $a_2 = -1$. In general, the helicity
modulus is given by the second derivative of the free energy with respect to the infinitesimal twist,
\begin{align}
\Upsilon_{\mu, (a_1, a_2)} &= \frac{1}{L^3}\frac{\partial^2
  F[\theta']}{\partial \delta_{\mu}^2}\Bigg|_{\bm{\delta} =
  \textbf{0}}\non\\
\begin{split}
&= \frac{1}{L^3}\Bigg[\bigg\langle \frac{\partial^2
      H[\theta']}{\partial \delta_{\mu}^2}\bigg \rangle-\beta\Bigg
    \langle \bigg(\frac{\partial H[\theta']}{\partial
        \delta_{\mu}}\\
&\quad-\bigg\langle\frac{\partial H[\theta']}{\partial
        \delta_{\mu}}\bigg\rangle\bigg)^2 \Bigg
\rangle\Bigg]\Bigg|_{\bm{\delta} = \textbf{0}},
\end{split}
\end{align}
where the notation $\theta'$ simply means that all phase variables are replaced according to Eq.
\eqref{Eq:twist}. In our case, with an isotropic system, we expect the helicity modulus to yield 
directionally independent results within statistical errors. For more details on the helicity modulus in the special case of the 
Villain model, we refer to Refs. \onlinecite{eskil1} and \onlinecite{Nguyen_98_1}.

\subsection{Gauge mass}
To capture the properties of the gauge field $\bfA$, we study the gauge-field correlator $\langle\bfA_{\bfq}\bfA_{-\bfq}\rangle$,
explicitly given for the lattice model, 
\be
\begin{split}
\langle \bfA_{\bfq}\bfA_{-\bfq}\rangle =&
\frac{2}{\beta(|\bfQ_{\bfq}|^2+m_{0}^{2})}\\
&\times\left[1+\frac{2\beta \pi^2 \mathcal{G}_{c,\bfq}}{|\bfQ_{\bfq}|^2(|\bfQ_{\bfq}|^2+m_{0}^{2})} \right],
\end{split}
\label{Eq:Gauge_field_correlator}
\ee
where $|\bfQ_{\bfq}|^2$ is the Fourier representation of the discrete
Laplace operator, given by $|\bfQ_{\bfq}|^2 =
\sum_{\mu}[2\sin(q_{\mu}/2)]^2$, with $q_{\mu} = 2\pi n_{\mu}/L$,
$n_{\mu} \in [1,...,L]$ and 
\begin{equation}
\mathcal{G}_{c,\bfq} = \left\langle e_je_lR_{jk}R_{lm}\bfm_{\bfq,k}\cdot\bfm_{-\bfq,m}\right\rangle,
\label{Vortex_correlator}
\end{equation}
is the correlation function of the linear combination of vortex fields
that corresponds to the charged sector of
Eq. \eqref{Eq:neutral_charged_vortex_modes}. Here $\bfm_{\bfq,j}$ is
the lattice model vortex field of component $j$ in Fourier space. The details of the derivation are given in Appendix \ref{App:gauge_field_corr}. 
In particular, we will use this quantity to extract the order parameter for the normal fluid-superconductor phase transition, 
i.e., the dynamically generated gauge-field mass or Higgs mass.  The
effective gauge mass $m_{\bfA}$ is extracted from the gauge-field
correlator by\cite{prb05, prl05-2, Hove_00_Kajantie_04} 
\begin{equation}
m_{\bfA}^2 = \lim_{\bfq = 0}\frac{2}{\beta \langle\bfA_{\bfq}\bfA_{-\bfq}\rangle}. 
\end{equation}
This quantity is employed as order parameter of 
the superconducting phase. Note that the dynamic creation of mass at $T_c$ and the onset of 
the Meissner phase, the manifestation of the Higgs mechanism in London superconductors, is 
governed entirely by the long-distance behavior of the vortex correlator of the charged mode, 
cf. Eqs. \eqref{Eq:Gauge_field_correlator} and \eqref{Vortex_correlator}.
In the ordered phase, where vortex loops are confined, 
$\lim_{\bfq \to 0} \langle \bfm_{k}(\bfq)  \bfm_{m}(-\bfq)\rangle \sim q^2$,
such that $\lim_{\bfq \to 0} \langle \bfA(\bfq)  \bfA(-\bfq)\rangle \sim const$, 
rendering the gauge field massive. When vortex loops proliferate, 
$\lim_{\bfq \to 0} \langle \bfm_{k}(\bfq)  \bfm_{m}(-\bfq)\rangle =$ const $\neq 0$, such that
$\lim_{\bfq \to 0} \langle \bfA(\bfq)  \bfA(-\bfq)\rangle \sim1/q^2$, rendering the
gauge field massless.   

In the Monte Carlo simulations, the vortex fields of both species are
extracted from the phase and gauge-field distributions by considering
the plaquette sum of the gauge-invariant phase difference,
\begin{equation}
\sum_{\nu \eta}\varepsilon_{\mu \nu \eta} \Delta_{\nu} (\Delta_{\eta}\theta_{\bfr, j} - e_j A_{\bfr, \eta}) = 2\pi m_{\bfr, \mu, j}, 
\end{equation}
where the left-hand side is the plaquette sum
of the gauge invariant phase difference, $\Delta_{\mu}\theta_{\bfr, j} - e_j A_{\bfr, \mu}$ and $m_{\bfr, \mu,
  j}$ is the real-space vortex field. The gauge invariant phase
difference must be kept in the primary interval for each link in the plaquette sum in 
order to accommodate vortices in the lattice model.
Now, by Fourier transformation of the vortex field,
$\mathcal{G}_{c,\bfq}$ is calculated, and to find the gauge mass,
curve fitting of the quantity 
$2/(\beta \langle\bfA_{\bfq}\bfA_{-\bfq}\rangle)$ is performed for
small $q$ values in order to extract the $q \rightarrow 0$ limit.

\section{Monte Carlo results, $\rho_d = 0, e_1 = e_2 = e$ \label{Sec:MCResults_no_drag}}
Here we present the simulation results for the case discussed in
Sec. \ref{Subsec:Easy_plane_case}. In this section we consider in general unequal
stiffnesses $\rho_1 \neq \rho_2$ in the regime where $\rho_d = 0$.

\begin{figure*}[tbp]
\includegraphics[width=0.95\textwidth]{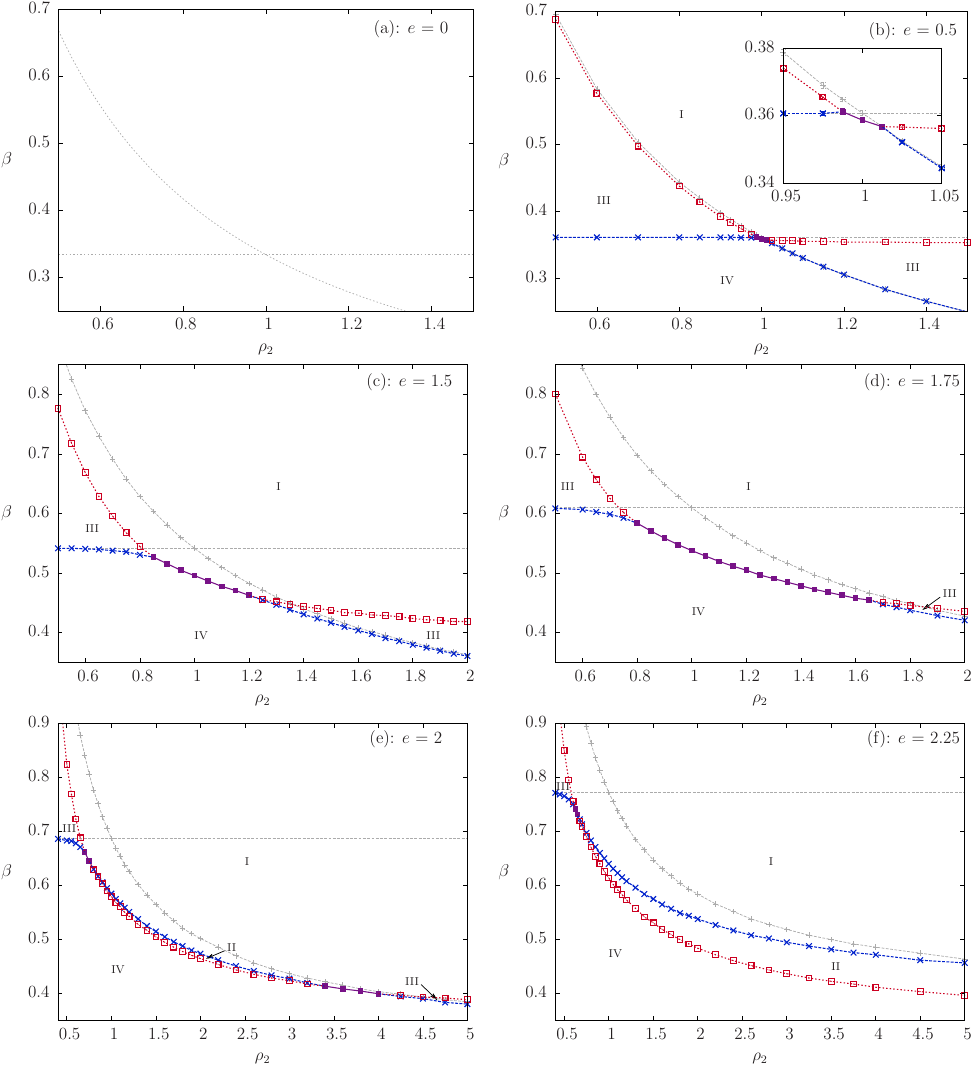}
\caption{(Color online) The phase diagram in the ($\rho_2$, $\beta$)-plane for 
the model in Eq. \eqref{gl2} at six increasing values of the electric
charge $e$ when $\rho_1 = 1$. Blue x
markers ($\times$) connected with dashed lines are charged critical points, and red squares ($\square$) connected with dotted lines are 
the neutral critical points. When
these critical points are merged, it is shown by filled
squares ($\blacksquare$) in violet connected with solid
lines. Moreover, with lines in silver color, we present
critical points of one-component superconductors with $e$ as denoted in
panel. The horizontal line is the critical line when $\rho = \rho_1 =
1$, and the plus markers ($+$) are the critical points when $\rho =
\rho_2$. For these reference lines, the dashed and dotted line type correspond to charged and neutral critical points, as above. 
The inset in panel (b) is a magnification
of the region where the lines merge. Phases are denoted
by roman numbers. I. Ordered phase with spontaneously broken $U(1)\times
U(1)$ symmetry, $m_{\bfA} \ne 0, \Upsilon_{\mu, (1, -1)} \ne
0$. II. Spontaneously broken global $U(1)$ symmetry, with restored
$U(1)$ gauge symmetry, $m_{\bfA} = 0, \Upsilon_{\mu, (1, -1)} \ne
0$. III. Spontaneously broken $U(1)$ gauge symmetry, with restored
global $U(1)$ symmetry, $m_{\bfA} \ne 0, \Upsilon_{\mu, (1, -1)} =
0$. IV. Normal phase with
fully restored $U(1)\times U(1)$ symmetry, $m_{\bfA} = 0, \Upsilon_{\mu, (1, -1)} =
0$. The system size considered is $32^3$. 
Except for inset, error bars are smaller than marker size and thus 
omitted from diagram.}
\label{Fig:Merge_tableaus}
\end{figure*}

Figure \ref{Fig:Merge_tableaus} shows the simulation results
varying the stiffness $\rho_2$, when the other stiffness $\rho_1$ is
set to unity. Results are obtained for six different values of the
electric charge, and we focus on the regimes where there is a strong
competition between proliferating topological defects. We locate the
critical inverse temperature of the charged and the neutral critical
point by locating the anomaly of the heat capacity associated with
the phase transition. The charged critical
point is associated with the point where the Meissner effect sets in,
evident by onset of the effective gauge mass $m_{\bfA}$, whereas the neutral critical
point is associated with the onset of the order in the gauge-invariant phase difference
with a corresponding nonzero value of the associated helicity
modulus $\Upsilon_{\mu, (1, -1)}$.

\subsection{Topological excitations}
Consider now the case when the neutral critical line is
situated above the charged critical line, that is, when going from
phase I  ($U(1)\times U(1)$ broken symmetry)
to phase III  (broken $U(1)$ charged symmetry)
across the neutral phase transition line in Fig. \ref{Fig:Merge_tableaus}.
This phase transition is driven either by proliferation of $(0,1)$  or $(1,0)$ vortices. 
The composite $(1,1)$ vortices do not couple to the neutral sector of Eq. \eqref{gl2} and
can thus never be responsible for destroying the order in the neutral sector. The
other composite topological excitation $(1,-1)$ is, by inspection
of Eq. \eqref{gl2}, seen to have neither energetic nor entropic advantage over individual
vortices. Because the vortices $(1,0)$, $(0,1)$ cost the same
amount of energy in the neutral sector, but the vortex with lowest
stiffness $\rho_j$ costs less energy in the charged sector,
 the neutral critical line must be associated with proliferation of individual vortices of the
component with the smallest value of the bare stiffness $\rho_j$, when
going from phase I to phase III. Outside the
region where there is a strong competition
between different kinds of vortex excitations, this phase transition is found to be of second order in the 3Dxy universality
class.\cite{prb05} When the individual vortices proliferate,
the corresponding stiffness is renormalized to zero and the remaining
condensate will be a charged condensate with order in the remaining component. 
Thus, the remaining condensate will, at a higher temperature, have a
phase transition similar to that of the following one-component superconductor,
\begin{equation}
F_{III\rightarrow IV}^{\textnormal{eff}} = \frac{\rho_j}{2}(\nabla\theta_j-e \bfA)^2 + \frac{(\nabla\times \bfA)^2}{2},
\label{Eq:1_comp_SC}
\end{equation}
where $j$ now is the index of the component with largest stiffness $\rho_j$. 

This is verified in Fig. \ref{Fig:Merge_tableaus} by
observing that the charged critical line between III and IV asymptotically approaches
the one-component reference lines away from the region
of competition between different kinds of topological
excitations. This phase transition is second order and in the 
universality class of the inverted 3Dxy transition.\cite{prb05}

When there is a transition from phase I
($U(1)\times U(1)$ broken symmetry)
to phase II (broken $U(1)$ neutral symmetry) in panels (e) and (f) of Fig. \ref{Fig:Merge_tableaus}, the charged critical point is
situated at a lower temperature than the neutral critical point. In
this case the topological defects responsible for
the phase transition are $(1,1)$ vortices, because the
other possible vortices will destroy order in the neutral sector of
Eq. \eqref{gl2}, and thus are not proliferating at this transition line.
  As discussed in Sec. \ref{Subsec:Easy_plane_case}, the $(1,1)$
vortices proliferating from an ordered background may be mapped onto a
single-component superconductor with effective stiffness (neglecting the 
internal structure of the vortices)
$\rho^{\prime} = \rho_{1}+\rho_{2}$, 
\begin{equation}
F_{I\rightarrow II}^{\textnormal{eff}} =
\frac{\rho_1+\rho_2}{2}(\nabla\theta-e\bfA)^2 + \frac{(\nabla\times \bfA)^2}{2}.
\label{Eq:I_to_II}
\end{equation}

\begin{figure*}[tbp]
\includegraphics[width=0.95\textwidth]{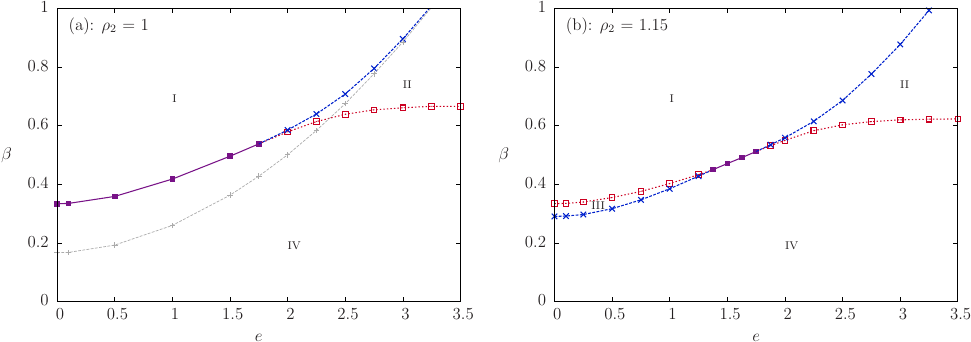}
\caption{(Color online) The phase diagram in the ($\beta$, $e$) plane for the two-component 3D London 
model, Eq. \eqref{gl2}, with $\rho_1 = 1$ and for two different values of
$\rho_2$. In the left diagram $\rho_2 = \rho_1 = \rho = 1$ whereas $\rho_2 = 1.15$ in the right
diagram, i.e., there is a moderate disparity. Markers and line types are the same as in Fig. \ref{Fig:Merge_tableaus}, 
i.e., blue x markers ($\times$) connected with dashed lines are charged critical points, red squares ($\square$) 
connected with dotted lines are 
neutral critical points, filled
squares ($\blacksquare$) in violet connected with solid
lines are merged transitions. The silvered plus markers ($+$) in the
left diagram is a one-component reference line of a superconductor
with bare stiffness $2\rho$ and charge $e$. Roman numbers denote the
different phases as given in the
caption of Fig. \ref{Fig:Merge_tableaus}. Note that these 
diagrams are 2D cross sections of a 3D phase diagram in ($\beta$,
$\rho_2$, $e$) space perpendicular to 
the cross sections in Fig. \ref{Fig:Merge_tableaus}. The lattice
size is $32^3$. Errors are smaller than marker size and thus
omitted from diagram.}  
\label{Fig:Eq_stiffnesses}
\end{figure*}

In Fig. \ref{Fig:Eq_stiffnesses} we show results when bare component
stiffnesses $\rho_j$ are kept fixed and electric charge $e$ is
varied. In panel (a), we also present a one-component reference line
corresponding to the phase transition  of the
superconductor in Eq. \eqref{Eq:I_to_II}. Indeed, away from the splitting point, the transition from I
to II approaches this reference line. Note that
the mapping in Eq. \eqref{Eq:I_to_II} yields a one-component
superconductor with stiffness $\rho_1 + \rho_2$ that always is stiffer
than the two reference lines in Fig. \ref{Fig:Merge_tableaus} (which
are one-component superconductors with stiffness $\rho_1$ and $\rho_2$). Thus, 
the charged transition line
between phase I and phase II is always lower than the reference
lines in Fig. \ref{Fig:Merge_tableaus}.
Phase II in
Figs. \ref{Fig:Merge_tableaus} and \ref{Fig:Eq_stiffnesses} is the metallic superfluid phase (i.e.,
exhibiting order only in the gauge-invariant phase difference) discussed in
Sec. \ref{Subsec:Easy_plane_case} and the effective free energy in the
remaining superfluid condensate is given in Eq. \eqref{glms}. The
cheapest topological defects that proliferate at higher temperatures  and destroy the
remaining composite order in this phase, are individual vortices.
Hence, away from the region of competing topological defects (i.e.,
away from the splitting point), the
transition line from phase II to phase IV 
should be similar to a one-component superfluid
with effective stiffness $\rho^{\prime} =
\rho_1\rho_2/(\rho_1+\rho_2)$, 
\begin{equation}
F_{II\rightarrow IV}^{\textnormal{eff}} = \frac{\rho_1\rho_2}{2(\rho_1+\rho_2)}(\nabla\theta)^2.
\label{Eq:II_to_IV2}
\end{equation}
Note that in both panels of Fig. \ref{Fig:Eq_stiffnesses}, the neutral transition line between II and IV is
found to be asymptotically independent of $e$, thus approaching a constant value asymptotically far away
from the region of competition with different vortices, as Eq. \eqref{Eq:II_to_IV2}
suggests. Moreover, Eq. \eqref{Eq:II_to_IV2} predicts the value $\beta
= \beta_c(\rho_1+\rho_2)/(\rho_1\rho_2)$ of the actual line, which
corresponds well with the results in the figure. Here, $\beta_c
\approx 0.334$ is the critical point of the one-component superfluid
($e = 0$) when $\rho = 1$.  

Note that vortices on the form $(n_1,n_2)$ with $n_{j} \geq 1$,
$n_{k\neq j} > 1$  can, by inspection of Eq. \eqref{gl2}, be shown to
 always be energetically unfavorable compared with
other topological excitations in this model. Such higher order
vortices are thus not relevant when $\rho_d = 0$ and $e_1 = e_2 = e$.

\subsection{Gauge-field fluctuation driven merger of the phase transitions in case of unequal bare stiffnesses}
We next discuss the evolution of the phase diagrams in
Figs. \ref{Fig:Merge_tableaus} and \ref{Fig:Eq_stiffnesses} when $e$ is 
varied. When charge increases, the energy of the composite $(1,1)$ vortices 
(which have no topological charge in the neutral sector), as well
as the energy associated with charged currents of individual vortices decrease.
This leads to a formation of a region in the phase diagram which is characterized by a \textit{merger 
of the two $U(1)$ transitions
in the case of unequal bare stiffnesses of the two condensates.}
Thus, even in the case
of unequal stiffnesses, when the coupling to a fluctuating
noncompact gauge field is sufficiently strong,
 there appears a phase
transition directly from the ordered phase with spontaneously broken
$U(1)\times U(1)$ symmetry to the fully disordered normal phase.
See also discussions of transition mergers caused by other kinds
of couplings in Refs. \onlinecite{kuklovAB,chernodub,eskil1}.
Panel (b) of Fig. \ref{Fig:Eq_stiffnesses} clearly illustrates this behavior. 
In this panel, the value of bare stiffness disparity is fixed when $e$ increases. For
low values of $e$ there are two phase transitions:
at lower temperature individual vortices with lower stiffness
proliferate while at higher temperature a proliferation
of individual vortices of stiffer condensate takes place.
However, when $e$ increases, the
two lines approach each other and merge at $e \approx 1.3$.

The line merger is a consequence of the fact that 
at a substantially large electric charge, the bare energy of an individual
 vortex in a broken $U(1)\times U(1)$ phase is dominated by the neutral mode.
 Because a proliferation 
of less energetically expensive individual defects destroys the neutral mode, 
this eliminates the bare long-range logarithmic interaction 
between vortices in the stiffer condensate, leading to a
 dramatic decrease in their bare line tension and thus
 to their {\it preemptive} proliferation.
 On the other hand in a range of parameters a
 proliferation of composite $(1,1)$ vortices 
 can trigger proliferation of individual vortices
 again leading to a  ``preemptive''
restoration of the full $U(1)\times U(1)$
symmetry via a single phase transition.
When electric charge is increased
further, then eventually at a certain point in the interval 
$ e \in (1.75 ... 1.875)$ the $(1,1)$ vortices become
much less energetically expensive than other excitations
and can proliferate at low temperatures
without triggering a
proliferation of individual vortices.
Then the metallic superfluid phase (II)
emerges as discussed in Sec. \ref{Subsec:Easy_plane_case}. 

\subsection{Order of the phase transition
associated with the merged lines  \label{Subsec:preemptive_transitions}}
Let us now characterize the phase transition
along the merged lines of Figs. \ref{Fig:Merge_tableaus} and
\ref{Fig:Eq_stiffnesses}. In Ref. \onlinecite{kuklov}, 
using the ${\bf j}$-current model 
the 
transition line from $U(1)\times U(1)$ to fully symmetric state
in the case of equal stiffnesses 
presented in panel (a) of
Fig. \ref{Fig:Eq_stiffnesses}, was found to be a
 first-order transition. We obtain consistent results in our
 Villain-model based simulations.

\begin{figure}[tb]
  \includegraphics[width=\columnwidth]{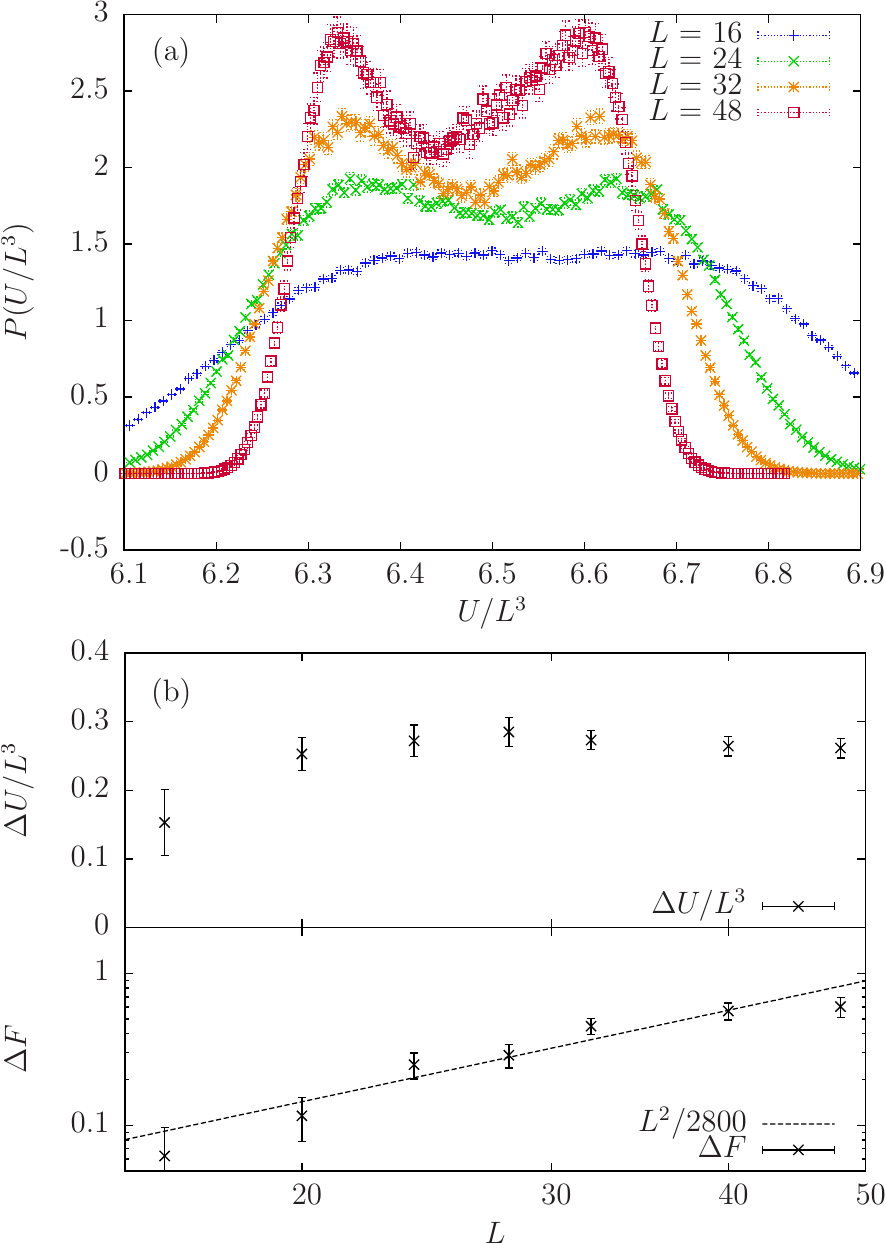}
  \caption{(Color online) (a) Histograms for the probability distribution
    of the internal energy per site $U/L^{3}$ at $\beta \approx 0.487$ when parameters 
    are $\rho_1 = 1$, $\rho_2 = 1.05$, and $e = 1.5$. This is the merged transition 
    point found in Fig. \ref{Fig:Merge_tableaus} at ($\rho_2$, $\beta$) = (1.05,
    0.487) in panel (c). A double-peak structure develops when $L$ increases. (b) Upper panel shows that the finite-size scaling of the 
    latent heat per site $\Delta U/L^{3}$ approaches a finite value when $L$ 
    increases. This is the distance between the peaks in (a). The lower panel shows 
    the finite-size scaling of the difference in the free energy, $\Delta F =
    (1/\beta)\ln(P_{\textnormal{max}}/P_{\textnormal{min}})$ taken between
    the double-peak value $P_{\textnormal{max}}$ and the value of the
    minimum in between $P_{\textnormal{min}}$ of the histograms in
    (a). For a first-order phase transition, $\Delta F \sim
    L^{d-1}$ (Ref. \onlinecite{Lee_90_91}). Ferrenberg-Swendsen reweighting was
    used to obtain histograms with similar height peaks (Ref. \onlinecite{Ferrenberg_88}).
  }
  \label{Fig:double_peak}
\end{figure}

Furthermore in Fig. \ref{Fig:double_peak}, we report the simulation
results associated
with the merged line
in a case when bare stiffnesses are not equal.
We find a first-order transition along the merged line in our case when there is a disparity of
the bare phase stiffnesses. This shows that the first-order phase 
transition in a $U(1) \times U(1)$ non-compact gauge theory is not related to the specific degeneracy of the model with equal stiffnesses 
$\rho_1 = \rho_2$ but appears to be related to the case when there are several competing 
or composite topological defects.

\section{Monte Carlo simulation, general model 
with both gauge field and dissipationless drag interactions \label{Sec:GeneralResults}}
Next, we present results from Monte Carlo simulations when both drag
and gauge field mediated interactions are included.

\subsection{Competing gauge field and drag interactions in the case
$\rho_1 = \rho_2 = 1$}
\begin{figure}[tb]
\includegraphics[width=\columnwidth]{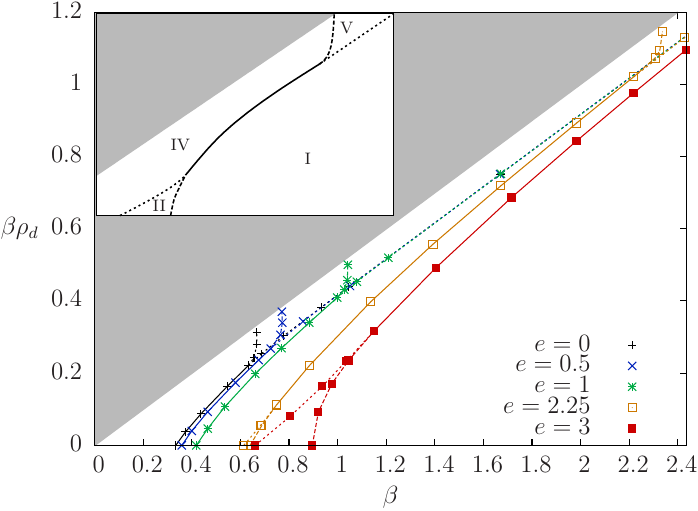}
\caption{(Color online) Phase diagram in the ($\beta$, $\beta\rho_d$)
  plane with competing gauge-field and drag interactions. Results are given for five different values of the charge
$e_1 = e_2 = e$. The bare component stiffnesses are equal, $\rho_1 = \rho_2 = 1$, and the 
system size considered is $32^3$. The gray-shaded area is the prohibited region by the stability 
condition, Eq. \eqref{Stability}. Line type corresponds to character of phase transition as in 
Fig. \ref{Fig:Merge_tableaus}, that is, charged lines are dashed, neutral lines are dotted, and 
merged transition lines are solid. Except for this, lines are guide to the eyes, only. The 
inset in the figure shows the schematic structure of the phases in the diagram for all cases with 
$e > 0$. Roman numbers denote phases. Phases I, II and IV are the same
as given in the caption of Fig. \ref{Fig:Merge_tableaus}, whereas V
is, similar to phase III in Fig. \ref{Fig:Merge_tableaus}, a phase
with spontaneously broken $U(1)$ gauge symmetry, and restored
global $U(1)$ symmetry, $m_{\bfA} \ne 0, \Upsilon_{\mu, (1, -1)} =
0$. However, in V the broken $U(1)$ gauge symmetry is associated with
composite phase sum, whereas in phase III of
Fig. \ref{Fig:Merge_tableaus}, it is associated with the phase of the single ordered component. 
For the given ranges of the phase diagram, II is only found for $e = 2.25$ and $e = 3$ and 
V is not found for $e = 3$. When $e = 0$, all phase transitions are neutral and phase II and
V are associated with broken global $U(1)$ symmetry in the phase difference and phase sum, 
respectively (Ref. \onlinecite{eskil1}). The results for $e = 0$ are here simulated with
a fluctuating gauge field and coincide (as they should) with the
equal stiffnesses results in Ref. \onlinecite{eskil1} with no fluctuating gauge field.}
  \label{Fig:PhaseDiagram_Villain}
\end{figure}

In Fig. \ref{Fig:PhaseDiagram_Villain}, we present results for the case
when the bare component stiffnesses are equal $\rho_1 = \rho_2 = \rho
= 1$,
and the gauge field couplings are equal, $e_1 = e_2 = e$. We vary the
inverse temperature $\beta$ and the bare drag coefficient $\rho_d$ and
map out the phase diagram in the $(\beta,\beta\rho_d)$ plane for a number of different values of $e$. We consider
positive $\rho_d$ only. In this specific case, the charged and neutral
modes in Eq. \eqref{Eq:Charged_and_neutral_modes} are written as 
\begin{equation}
\begin{split} 
F =
\frac{1}{2}\Bigg[&\frac{\rho-2\rho_d}{2}(\nabla
\theta_1-\nabla \theta_2)^2\\
&+ \frac{\rho}{2}(\nabla
  \theta_1 + \nabla \theta_2 - 2e{\bf A})^2 + (\nabla \times {\bf A})^2 \Bigg].
\end{split}
\label{Eq:Charged_and_neutral_modes_spec_case_1}
\end{equation} 

Here, we have the interesting situation where drag- and gauge-field mediated intercomponent long-range vortex
interactions are found to be of opposite signs, see Eq. \eqref{Eq:Dual_V}. Thus,  the drag coupling
$\rho_d \ne 0$, when significantly strong, favors formation of the 
 $(1,-1)$ composite vortices (via a mechanism similar to that in Ref. \onlinecite{eskil1}).
On the other hand, the gauge field coupling favors
the formation of $(1,1)$ bound states of individual vortices when $e_1$ and $e_2$
are of the same sign. This competition is studied in Fig. \ref{Fig:PhaseDiagram_Villain}.
Its most striking consequence  is that it leads to the existence
of four phases: at strong drag there is a superconducting phase where a neutral
mode is destroyed by the proliferated $(1,-1)$ 
vortices  (phase V). At strong electric charge
there is a superfluid phase with proliferated $(1,1)$ vortices (phase II). 

We next consider these phases more closely. The results in Fig. \ref{Fig:PhaseDiagram_Villain}
show that the phase V appears when  $(1,-1)$ vortices proliferate
and thus there is no longer a broken symmetry in the neutral sector of
Eq. \eqref{Eq:Charged_and_neutral_modes_spec_case_1}. Note that when
we are well above the region of competing topological defects in Fig. 
\ref{Fig:PhaseDiagram_Villain}, then, by
neglecting the internal vortex structure, we may 
approximate the $(1,-1)$ vortices to map 
onto vortices in a one-component superfluid with stiffness $\rho^{\prime} = 2(\rho-2\rho_d)$, 
\begin{equation}
F_{I\rightarrow V}^{\textnormal{eff}} = (\rho-2\rho_d)(\nabla\theta)^2.
\label{Eq:I_to_V_gen_case}
\end{equation}
This effective limiting model is $e$ independent. Indeed
this physics manifests itself in the fact that in
Fig. \ref{Fig:PhaseDiagram_Villain}, the actual transition is seen to
approach asymptotically the reference line $\beta\rho_d = (\beta-\beta_c/2)/2$.

In Sec. \ref{Sec:MCResults_no_drag}, the superconducting phase III,
which similarly to phase V exhibits order in the charged sector and
disorder in the neutral sector, was
created from the fully ordered phase by proliferation of
individual vortices when we increased disparity
in the bare stiffness of the two components. Here, phase V is
created by proliferation of composite vortices and the coupling constant
responsible for creating the phase is $\rho_d$. Consequently, the remaining order is now in the gauge-invariant
phase difference of the charged mode, given by second and third terms
in Eq. \eqref{Eq:Charged_and_neutral_modes_spec_case_1}. On the other hand, phase III exhibits order
in the component with largest
bare stiffness. Note also that in the $U(1)\times U(1)$ (phase I) state with equal stiffnesses the  $(1,0)$, $(0,1)$ vortices
carry half of the magnetic flux quanta.\cite{prl02} It can be seen
from Eq. \eqref{Eq:Charged_and_neutral_modes_spec_case_1} that in the phase V $(1,0)$, $(0,1)$ vortices
become equivalent and no longer have logarithmic divergence of internal energy per unit length 
due to absence of a neutral mode. That is, they become similar to 
Abrikosov vortices, but carry only a half quantum of magnetic
flux. This phenomenon is related to the fractionalization of superfluid
velocity quantum in the metallic superfluid state.\cite{nphys} From
Eq. \eqref{Eq:Charged_and_neutral_modes_spec_case_1} 
it also follows that the individual vortices
behave as vortices in a one-component superconductor with effective
stiffness $\rho^{\prime} = \rho/2$ and double effective charge $e^{\prime} = 2e$,
\begin{equation}
F_{V\rightarrow IV}^{\textnormal{eff}} = \frac{\rho}{4}(\nabla\theta-2e \bfA)^2 + \frac{(\nabla\times \bfA)^2}{2}.
\label{Eq:1_comp_SC_gen_case}
\end{equation}
In Fig. \ref{Fig:PhaseDiagram_Villain}, the transition from the phase V to the normal phase IV is indeed found 
to tend asymptotically to a phase transition one would predict from
the model, Eq. (\ref{Eq:1_comp_SC_gen_case}). For 
this model, the transition line is found to be vertical, in accordance with the drag independent stiffness in 
Eq. \eqref{Eq:1_comp_SC_gen_case}. Note that when $e$ increases, the critical temperature of the vortex loop 
proliferation is decreased and the vertical line moves to the right in Fig. \ref{Fig:PhaseDiagram_Villain}. 

Next, the phase II may be investigated in a similar way as the phase V above. Phase II appears when $(1,1)$ vortices 
proliferate. As discussed in Sec. \ref{Sec:MCResults_no_drag}, the remaining order is in the neutral sector 
of Eq. \eqref{Eq:Charged_and_neutral_modes} and the transition to the normal state is governed by proliferation of 
individual vortices that asymptotically behave as a one-component superfluid with effective stiffness 
$\rho^{\prime} = (\rho-2\rho_d)/2$, 
\begin{equation}
F_{II\rightarrow IV}^{\textnormal{eff}} = \frac{\rho-2\rho_d}{4}(\nabla\theta)^2.
\label{Eq:II_to_IV_gen_case}
\end{equation}
The phase transition of this condensate will follow the line $\beta\rho_d =
(\beta-2\beta_c)/2$. Indeed, this is the case for $e = 3$ in Fig. \ref{Fig:PhaseDiagram_Villain} 
away from the region with competing topological defects.

\begin{figure}[tb]
  \includegraphics[width=\columnwidth]{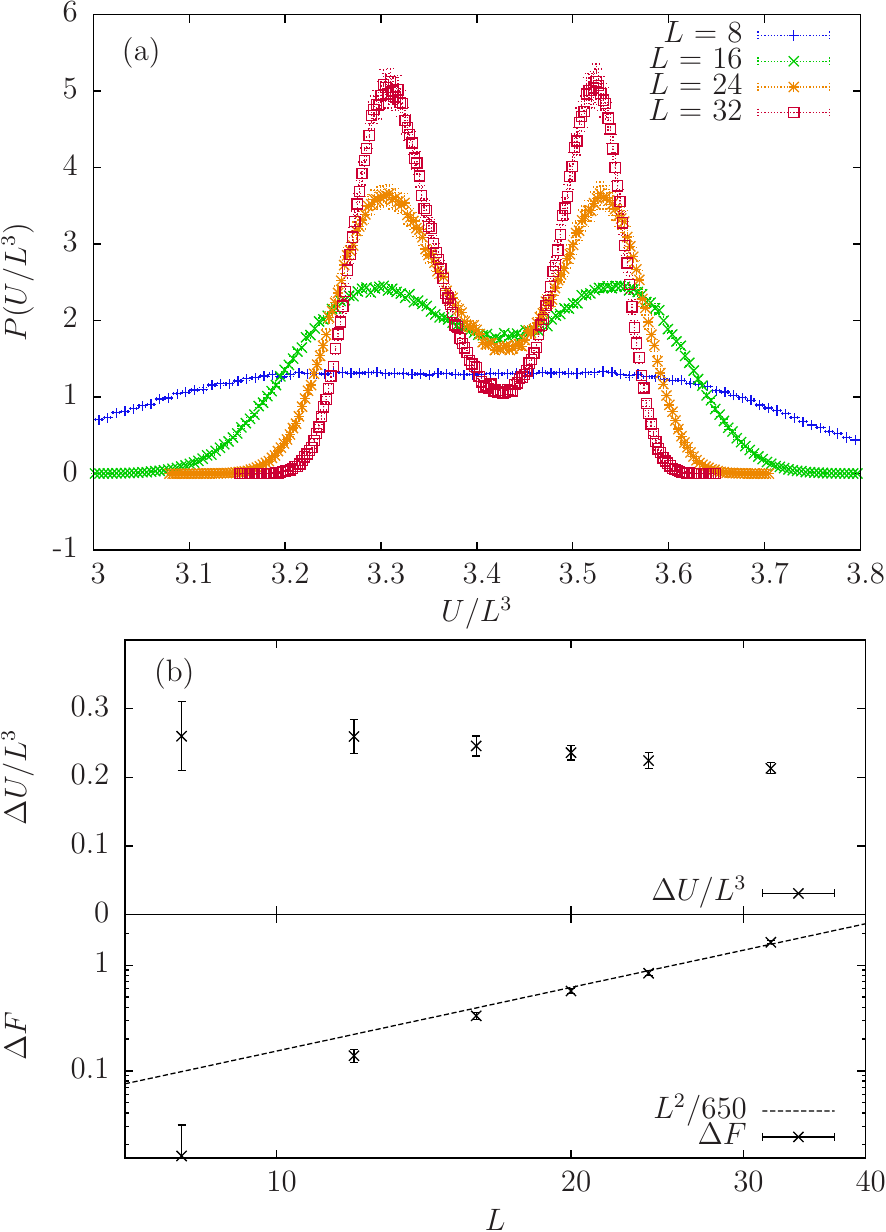}
  \caption{(Color online) (a) Histograms for the probability distribution
    of the internal energy per site $U/L^{3}$ at the critical point when
    parameters  are $\rho_1 = \rho_2 = \rho = 1$, $e_1 = e_2 = e =
    1$, and $\rho_d = 0.4$. This is a merged transition point at ($\beta$, $\beta
    \rho_d$) = (0.948, 0.379) along the critical line for $e = 1$ in
    Fig. \ref{Fig:PhaseDiagram_Villain}. A pronounced double-peak structure is
    found to develop when $L$ increases. (b) Upper panel shows the
    finite-size scaling of the latent heat per site $\Delta U/L^{3}$. This is the distance between the peaks in
    (a). The lower panel shows the finite-size scaling of the difference in the free energy,
    $\Delta F =
    (1/\beta)\ln(P_{\textnormal{max}}/P_{\textnormal{min}})$ taken between
    the double-peak value $P_{\textnormal{max}}$ and the value of the
    minimum in between $P_{\textnormal{min}}$ of the histograms in
    (a). For a first-order phase transition, $\Delta F \sim
    L^{d-1}$ (Ref. \onlinecite{Lee_90_91}). Ferrenberg-Swendsen reweighting was
    used to obtain histograms with peaks of similar height (Ref. \onlinecite{Ferrenberg_88}).
  }
  \label{Fig:double_peak_gen}
\end{figure}

Similarly to Sec. \ref{Sec:MCResults_no_drag} we find evidence of a
first order transition when lines are merged and $e > 0$, as seen in
Fig. \ref{Fig:double_peak_gen}. When only drag or gauge field is
included in a two-component system, first-order transitions may
emerge.\cite{eskil1,kuklov,steinar} Our results show that the
first-order character of this phase-transition line persists also in the 
case where both of the interactions are present and competing. 

\subsection{Regime where gauge field and drag interactions 
both favor formation of similar paired phase}
\begin{figure}[tb]
  \includegraphics[width=\columnwidth]{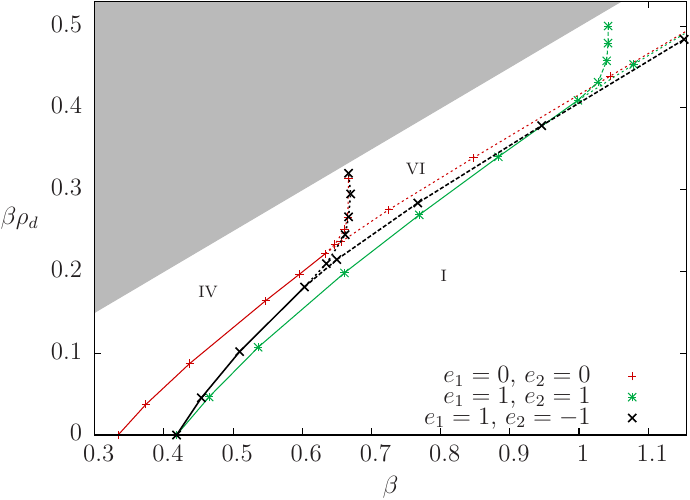}
  \caption{Phase diagram in the ($\beta$, $\beta\rho_d$) plane of the general
  model when $\rho_1 = \rho_2 = 1$ for the case of different charges
  $e_1 = -e_2 = 1$. These are the black x markers ($\times$) and the line type 
  denotes charged, neutral, and merged critical lines by dashed, dotted, and solid 
  lines as in Figs. \ref{Fig:Merge_tableaus}, \ref{Fig:Eq_stiffnesses}, and
  \ref{Fig:PhaseDiagram_Villain}. Roman numbers denote the phases of
  this particular case, $e_1 = -e_2 = 1$. I. Ordered phase with spontaneously 
  broken $U(1)\times
  U(1)$ symmetry, $m_{\bfA} \ne 0, \Upsilon_{\mu, (1, 1)} \ne
  0$. IV. Normal phase with
  fully restored $U(1)\times U(1)$ symmetry, $m_{\bfA} = 0, \Upsilon_{\mu, (1, 1)} =
  0$. VI. Spontaneously broken global $U(1)$ symmetry, with restored
  $U(1)$ gauge symmetry, $m_{\bfA} = 0, \Upsilon_{\mu, (1, 1)} \ne
  0$. For comparison, the results of the two cases $e_1 = e_2 =
  0$, $1$, from Fig. \ref{Fig:PhaseDiagram_Villain}, are presented. The phases for 
  these two cases follow from the inset and caption of Fig. \ref{Fig:PhaseDiagram_Villain}. 
  The lines are guide to the eyes. The system size considered is $32^3$. The uncertainties 
  in the position of the phase-transition lines are smaller than
  the marker size and are omitted from the diagram.} 
  \label{Fig:Opposite_Charges}
\end{figure}

In Fig. \ref{Fig:Opposite_Charges} we present the phase diagram in the
case when $\rho_1 = \rho_2 = \rho = 1$ and $e_1 = -e_2 = e = 1$. The
separation in neutral and charged modes is now, 
\begin{equation}
\begin{split} 
F =
\frac{1}{2}\Bigg[&\frac{\rho-2\rho_d}{2}(\nabla
  \theta_1 - \nabla \theta_2 - 2e{\bf A})^2 + (\nabla \times {\bf A})^2\\
&+ \frac{\rho}{2}(\nabla
\theta_1+\nabla \theta_2)^2 \Bigg].
\end{split}
\label{Eq:Charged_and_neutral_modes_spec_case_2}
\end{equation} 
The motivation
for investigating this particular case is found in the off-diagonal elements of 
the matrix in Eq. \eqref{Eq:Dual_V} where the interactions originating with the
gauge field will act in unison with the bare drag interactions upon
switching the sign of the electric charge in one of the
components (in contrast to the situation
considered in the previous section). 

Consider the simulation results shown in
Fig. \ref{Fig:Opposite_Charges}.
For comparison, we include the results when
there is no gauge-field coupling, $e_1 = e_2= 0$, and when gauge-field coupling competes
with the drag interaction, $e_1 = e_2= 1$. First notice that the
paired phase which appears when charges are opposite, 
is the metallic superfluid phase (VI) which now is associated with spontaneously broken global $U(1)$
symmetry in the phase sum (and not the phase difference as in
Figs. \ref{Fig:Merge_tableaus}, \ref{Fig:Eq_stiffnesses}, and \ref{Fig:PhaseDiagram_Villain}). Positive drag will favor $(1,-1)$ vortices as before. 
However, because of the change in sign of one of the charges, the $(1,-1)$ vortices are now associated 
with the charged sector of
Eq. \eqref{Eq:Charged_and_neutral_modes_spec_case_2}. The $(1,1)$
vortices are associated with the neutral mode, and thus the neutral critical point is determined by the 
onset of the associated helicity modulus $\Upsilon_{\mu, (1, 1)}$. The gauge-field renders the $(1,-1)$ 
vortices the topological objects with lowest excitation energy. When they proliferate the superconducting 
sector is destroyed.  Asymptotically, the associated phase-transition line is therefore expected to follow 
the behavior of a one-component superconductor with $\rho^{\prime} = 2(\rho-2\rho_d)$ and effective charge $e$,
\begin{equation}
F_{I\rightarrow VI}^{\textnormal{eff}} =
(\rho-2\rho_d)(\nabla\theta-e\bfA)^2 + \frac{(\nabla\times \bfA)^2}{2}.
\label{Eq:I_to_II_opposite_case}
\end{equation}
The remaining condensate will have superfluidity destroyed via proliferation
of individual vortices which asymptotically can be mapped
onto a one-component superfluid with stiffness
$\rho^{\prime} = \rho/2$,
\begin{equation}
F_{VI\rightarrow IV}^{\textnormal{eff}} = \frac{\rho}{4}(\nabla\theta)^2.
\label{Eq:II_to_IV_opposite_case}
\end{equation}
This is exactly the same behavior as expected when $e = 0$, 
which is also confirmed by simulations in
Fig. \ref{Fig:Opposite_Charges}. Note that there is neither $\rho_d$ nor
$e$ dependence of this line.

Figure \ref{Fig:Opposite_Charges} also shows that when gauge-field and drag act in unison it amounts to a small 
increase in the region of paired  phase compared to the case
when there is only drag interaction. However when interactions compete there is 
a stronger effect of the suppression of the corresponding paired
phase. Also note that the cases $e_1 = -e_2 = 1$ and 
$e_1 = e_2 = 1$ coincide when $\rho_d = 0$. This is readily inferred from Eq. \eqref{Eq:london}, when $\rho_d=0$, one model can be mapped onto another
by change in sign of charges $e_j$ accompanied by a sign change
in one of the phases $\nabla \theta_j$. 

\section{$N$-component case \label{Sec:N-comp_case}}
In the case of $N$ charged components, the $[U(1)]^N$ model can be written as (see Appendix A for notation)
\begin{eqnarray}
 F &=& \sum_{ j =1,...N} \frac{\rho_j}{2}(\nabla  \theta_j-e_j \bfA)^2 + \frac{(\nabla\times \bfA)^2}{2} 
 \nonumber \\
 &-& \sum_{ j,k=1,...N}\frac{\rho_{d,jk}}{2}  (\nabla \theta_j- e_j \bfA - \nabla \theta_k + e_k \bfA)^2.
 \label{Eq:london2}
\end{eqnarray}
In the $N$-component case the phase structure becomes more complex than in the two-component case.
In the simplest case of different stiffnesses and weak coupling,
there  can take place a vortex proliferation 
in individual fields. That  reduces the symmetry to the $[U(1)]^{N-1}$.
On the other hand, in the case when the gauge-field coupling is dominant there can take
place a proliferation of one-flux-quanta composite vortices while
individual vortices remain confined.
Consider the case of all equal charges. Then, in the $N$-component model
such a vortex has the phase winding $(\Delta \theta_1=2\pi,\Delta \theta_2=2\pi,...,\Delta \theta_N=2\pi)$.
When charge is sufficiently large, such a composite object can proliferate while the other
kind of (fractional flux) vortices remain confined.
The resulting state has broken global symmetries associated 
with all the combinations of the gauge-invariant phase differences $(\theta_j-\theta_k)$.
This is the $N$-component analog of the metallic superfluid state which has no Meissner
effect because of restored symmetry in the charged sector.

On the other hand, by varying the intercomponent drag strength along with the charge strength 
one can make different topological excitations the energetically cheapest
objects (such as, e.g., $(\Delta \theta_1=2\pi M_1,\Delta \theta_2=2\pi M_2, ... ,\Delta \theta_N=2\pi M_N)$,  
with any integers $M_N=0,\pm 1, \pm 2, ...,$ etc.).
This also can be seen from the intervortex interactions derived in Appendix \ref{App:N_comp_dualization}.
Proliferation of such objects reduces  broken $[U(1)]^{N}$ symmetry down to broken symmetries associated with various
 {\it weighted }combinations of phases such as $(L_1 \theta_1 + L_2 \theta_2 + L_3 \theta_3 + ...)$.
Consider, for example, the case where all kinds of ``two-vortex'' bound states proliferate, i.e.,
when the drag coupling makes the following objects energetically cheapest to excite,
$(\Delta \theta_1=2\pi,\Delta \theta_2=-2\pi,\Delta \theta_3=0, ... ,\Delta \theta_N=0)$,
$(\Delta \theta_1=0,\Delta \theta_2=2\pi,\Delta \theta_3=-2\pi,\Delta
\theta_4=0, ... ,\Delta \theta_N=0 )$,...,$(\Delta \theta_1=0,...,\Delta \theta_{N-2}=0,\Delta \theta_{N-1}=2\pi,\Delta \theta_N=-2\pi)$.
When these kind of topological defects (with two opposite phase
windings in different phases) proliferate, the only remaining broken symmetry is
associated with the sum of all phases $(\theta_1+\theta_2+...\theta_N)$ yielding the effective model,
\begin{equation}
\frac{1}{2} \rho_{\rm eff}[\nabla (\theta_1+\theta_2+...+\theta_N) - Ne\bfA]^2+ \frac{1}{2} (\nabla \times \bfA)^2,
\end{equation}
the prefactor $N$ in front of the vector potential $\bfA$ means that this is 
a ``charge-$Ne$'' superconductor, i.e., in this state only co-directed 
electrical current of all components is dissipationless.

\section{Conclusion \label{Sec:Conclusion}}
We have considered a three dimensional lattice superconductor model in
the London limit,
with two and $N$ individually conserved condensates. These condensates
interact with each other by two
mechanisms. The first is a dissipationless Andreev-Bashkin drag term
representing a current-current
interaction. The second is a fluctuating gauge-field. Intercomponent
Josephson coupling
is absent on symmetry grounds.
Such models are relevant in a number of physical circumstances ranging from the 
theories of the quantum ordered states of metallic hydrogen, models of neutron stars,
and were earlier suggested as effective models describing
valence-bond solid to Neel quantum phase transition in the proposed theories of deconfined quantum criticality.

In the $U(1) \times U(1)$ case when there is no intercomponent drag, $\rho_d = 0$, and component
charges are equal, $e_1 = e_2 = e$ and there is a disparity of the
bare component stiffnesses, we find that a sufficiently strong coupling to a non-compact 
gauge-field causes a merger of phase-transition lines. This  yields a direct transition from
broken $U(1)\times U(1)$ to normal state even when the bare component
stiffnesses are unequal.
When the charge $e$ is increased, the  merger
occurs for a higher disparity of stiffnesses.
However,  a further increase in the coupling beyond a
certain critical strength results in a new splitting of
the transition line. This yields a metallic superfluid phase. The
merger of the $U(1)$ transition lines is associated with a competition
between different kinds of
topological defects where proliferation of one type of vortices triggers a
preemptive proliferation of another. The result is a  much more complex
picture of the behavior
of topological defects in the  phase transition  than in
single-component $U(1)$ models.
The second splitting is due to the fact that increased coupling 
to the non-compact gauge-field decreases the  free energy per unit length
of a  bound state of topological defects. The bound state in
question (a composite vortex) has a topological charge
only in the charged sector of the model. This in turn
results in increased suppression of the critical stiffness associated
with the  charged sector of the theory, which eventually undergoes a
symmetry-restoring phase transition before the neutral sector.

We find that also when the bare stiffnesses
are unequal, the merged phase transition is first order in character.
Note that previously first-order transitions were reported in the
$U(1) \times U(1)$  gauge theory with degenerate stiffnesses,\cite{kuklov,steinar} $U(1) \times U(1)$ models with a compact
gauge field, as well as to phase transitions
in the SU(2) model with noncompact Abelian gauge-field.
\cite{chernodub}

In the main part of the paper, we have performed
a study of the phase diagram of the
generic $U(1) \times U(1)$ lattice London gauge model
featuring  both gauge-field and direct non-dissipative drag interactions. We
have
obtained, through large-scale Monte Carlo simulations, its phase diagram
as a function of these two generic coupling constants.

For the case where the bare component stiffnesses and charges are equal,
$\rho_1 = \rho_2 = 1$ and $e_1 = e_2 = e$,  we find the formation of two
different paired phases as a result of a competition between gauge-field and
intercomponent drag couplings. High values of drag produce a composite superconducting phase
associated with a broken local $U(1)$ gauge symmetry in the phase sum.
 There, the theory effectively features a doubled electric charge compared with
$U(1)\times U(1)$ phase, cf. Eq. \eqref{Eq:1_comp_SC_gen_case}.
At high values of $e$, the  gauge-field coupling wins over the drag coupling,
yielding  a paired superfluid phase (the  metallic superfluid) associated with
the
order in the gauge invariant phase difference. In between these two different
phases,
there is a region with a direct transition from broken $U(1)\times U(1)$ to
normal
state, part of which exhibits clear-cut signatures of a first-order transition, cf.  the
transition
line connecting regions II and V in Fig. \ref{Fig:PhaseDiagram_Villain}.

For comparison, we reported a quantitative study of the situation where
gauge-field mediated intercomponent interactions and intercomponent
drag both favor metallic superfluid phase.

In the final part of the paper we discussed the physics 
of states with composite symmetry breakdowns in the $N$-component 
London superconductor.

\acknowledgments  
We acknowledge useful discussions with I. B. Sperstad and E. B. Stiansen.  
E.V.H. thanks NTNU for financial support. E.B. was supported by Knut and Alice Wallenberg 
Foundation through the Royal Swedish Academy of Sciences Fellowship,
Swedish Research Council and by the National Science Foundation CAREER Award No. DMR-0955902.
A. S. was supported 
by the Norwegian Research Council under Grant No. 167498/V30  (STORFORSK).
E.B.  and A.S.  acknowledge the hospitality of the Aspen Center for Physics, where part of this work was done.

\appendix

\section{Vortex interaction in the $N$-component model \label{App:N_comp_dualization}}
{In the case of  arbitrary number of components $N$  the action has the form}

\begin{gather}
Z = \int \mathcal{D}\theta_1 ... \int \mathcal{D}\theta_N \int \mathcal{D}
\bfA \quad \e^{-S},\non
\\ 
\begin{split}
S &= \frac{\beta}{2}\int d^3r \Big\{[\nabla \times \bfA(\bfr)]^2\\
&\quad + [\nabla\theta_j(\bfr)-e_j \bfA(\bfr)]R_{jk}[\nabla\theta_k(\bfr)-e_k \bfA(\bfr)]\Big\}.
\label{Eq:General_N}
\end{split}
\end{gather}
The matrix $R_{jk}$ is in general given by 
\begin{equation}
R_{jk} = \left(\rho_j - \sum_{l}\rho_{d, jl} \right)
\delta_{jk} + \rho_{d, jk},
\end{equation}
where $\rho_{d, jk}$ is the drag coefficient between components $j$
and $k$, obviously, $\rho_{d, jk} = \rho_{d, kj}$ and $\rho_{d,
  jk} = 0$ when $j = k$. Following exactly the same procedure as in
the case $N=2$, we arrive at the $N$-component action,
 
\begin{equation} 
\begin{split}
S = \frac{\beta}{2} \int d^3q &\Bigg[\textbf{U}_j(\bfq)\left(R_{jk}-\frac{e_le_mR_{lj}R_{mk}}{q^2+m_{0}^{2}}\right)\textbf{U}_k(-\bfq)\\
&\quad+ \tilde{\bfA}(\bfq)(q^2+m_{0}^{2})\tilde{\bfA}(-\bfq)\Bigg], 
\end{split}
\end{equation}
where the Fourier transform of $\nabla \theta_j(\bfr)$, is denoted by
$\textbf{U}_j(\bfq)$, and
\begin{equation}
m_{0}^{2} = e_jR_{jk}e_k. \label{Eq.general_m_0}
\end{equation}
This expression is seen to reproduce the case $N=2$ given in Eq. \eqref{Eq:M}. The gauge-field is  
integrated out and the dualization now follows the same path as previously, yielding
\begin{gather}
Z = \sum_{\bfm_1}...\sum_{\bfm_N} \quad \e^{-S},\non\\
S = 2\beta \pi^2 \int d^3q \; \bfm_j(\bfq)V_{jk}(q^2)\bfm_k(-\bfq), 
\label{Eq:General_N_Dual_S}
\end{gather}
where the vortex interactions are given by
\begin{align}
V_{jk}(q^2) &= \frac{1}{q^2}  \left( R_{jk} - \frac{e_le_mR_{lj}R_{mk}}{q^2+m_{0}^{2}} \right) \non\\
           &= \frac{R_{jk} - \frac{e_le_mR_{lj}R_{mk}}{m_{0}^{2}}}{q^2} +
\frac{\frac{e_le_mR_{lj}R_{mk}}{m_{0}^{2}}}{q^2 + m_{0}^{2}}.
\label{Eq:General_N_Vortex_int}
\end{align}
This is seen to be on precisely the same form as Eq. \eqref{Eq:Dual_V}
for the case $N=2$. Following Appendix B in
Ref. \onlinecite{prb05}, a dualization of the corresponding
two-component lattice model in Eq. \eqref{Eq:Villain_H} may be performed to
yield the exact same result as in Eqs. \eqref{Eq:General_N_Dual_S} and \eqref{Eq:General_N_Vortex_int} where the vortex
fields now are defined on the vertices of the Fourier space dual
lattice and $q^2 \rightarrow |\bfQ_{\bfq}|^2$.

\section{Gauge-field correlator}
\label{App:gauge_field_corr}
By adding source term and Fourier transformation of the model in
Eq. \eqref{Eq:General_N}, the generating functional for deriving the
gauge-field correlator reads

\begin{gather}
Z_J = \int \mathcal{D}\theta_1 ... \int \mathcal{D}\theta_N \int \mathcal{D}
\bfA \quad \e^{-S},\non
\\ 
\begin{split}
S_J =& \frac{\beta}{2}\int d^3q
\Big\{\frac{1}{\beta}\left[\bfJ(\bfq)\bfA(-\bfq)+\bfA(\bfq)\bfJ(-\bfq)
\right]\\
&+
[\textbf{U}_j(\bfq)-e_{j}\bfA(\bfq)]R_{jk}[\textbf{U}_k(-\bfq)-e_{k}\bfA(-\bfq)]\\
&+q^{2}\bfA(\bfq)\bfA(-\bfq)\Big\},
\end{split}
\end{gather}
where $\bfJ(\bfq)$ are the electric currents that couples
linearly to the gauge-field in the source terms. Sum over repeated indices is assumed. We now proceed
similar to Sec. \ref{Subsec:Dual_model} by completing the squares of
the gauge-field and integrate out the shifted gauge-field
$\tilde{\bfA}(\bfq) =
\bfA(\bfq)+(\bfJ(\bfq)/\beta-e_{j}R_{jk}\textbf{U}_k(\bfq))/(q^2+m_{0}^{2})$
which yields 
\begin{gather}
Z_J = \int \mathcal{D}\theta_1 ... \int \mathcal{D}\theta_N \quad \e^{-S},\non
\\ 
\begin{split}
S_J =& \frac{\beta}{2}\int d^3q
\Bigg\{-\frac{\bfJ(\bfq)\bfJ(-\bfq)}{\beta^{2}(q^2+m_{0}^{2})}\\
&+ \frac{\bfJ(\bfq)e_{j}R_{jk}\textbf{U}_k(-\bfq)+e_{j}R_{jk}\textbf{U}_k(\bfq)\bfJ(-\bfq)}{\beta(q^2+m_{0}^{2})}\\
&+\textbf{U}_j(\bfq)\left(R_{jk}-\frac{e_le_mR_{lj}R_{mk}}{q^2+m_{0}^{2}}\right)\textbf{U}_k(-\bfq)\Bigg\}.
\label{Eq:GeneratingFunctional}
\end{split}
\end{gather}

We now employ the constraint $\nabla\cdot\bfJ(\bfr) = 0$,
i.e., the electrical currents are divergence free, such that components
parallel to $\bfq$ are unphysical. Thus, the physical
components of $\bfJ(\bfq)$ in the first term of Eq. \eqref{Eq:GeneratingFunctional}
are projected out with the transverse projection operator,
\begin{equation}
P_{T, \mu \nu} = \delta_{\mu \nu} - \frac{q_{\mu}q_{\nu}}{q^2}.
\end{equation}
As discussed in Sec. \ref{Subsec:Dual_model}, we disregard the
longitudinal part of $\textbf{U}_j(\bfq)$ and introduce the Fourier
transformed vortex fields by Eq.
\eqref{Eq:Transversal_part_fourier_rep}. Thus, the generating
functional is written as

\begin{gather}
Z_J = \sum_{\bfm_1}...\sum_{\bfm_N} \quad \e^{-S_{0}-S_{1}},\non\\
S_{0} = 2\beta \pi^2 \int d^3q \;
\bfm_j(\bfq)V_{jk}(q^2)\bfm_k(-\bfq),\non\\
\begin{split}
S_{1} = \int d^3q&\Bigg\{\frac{i\pi e_{j}R_{jk}\varepsilon_{\mu \nu
    \eta}q_{\nu}}{q^2(q^2+m_{0}^{2})}\\ 
&\;\times\left[m_{\eta, k}(\bfq)
  J_{\mu}(-\bfq) - J_{\mu}(\bfq) m_{\eta, k}(-\bfq) \right]\\
&- \frac{J_{\mu}(\bfq)P_{T, \mu \nu}J_{\nu}(-\bfq)}{2\beta(q^2+m_{0}^{2})}\Bigg\},
\end{split}
\end{gather}
where $V_{jk}(q^2)$ is given by Eq. \eqref{Eq:General_N_Vortex_int} and
$\varepsilon_{\mu \nu \eta}$ is the Levi-Civita symbol. Note that
there is an implicit sum over all indices $j$, $k$, $\mu$, $\nu$,
and $\eta$.
 
The gauge-field correlators are derived the standard way by
functional derivation of the currents,
\begin{align}
\langle A_{\mu}(\bfq) A_{\nu}(-\bfq)\rangle &=
\frac{1}{Z_0}\frac{\delta^2 Z_J}{\delta J_{\mu}(-\bfq) \delta J_{\nu}(\bfq)}\Bigg|_{\bfJ = \textbf{0}}\non\\
&= \left \langle \frac{\delta^2\e^{-S_1}}{\delta J_{\mu}(-\bfq) \delta J_{\nu}(\bfq)}\Bigg|_{\bfJ = \textbf{0}}\right \rangle,
\label{Eq:prop_definition}
\end{align}
where $Z_0 = Z_J|_{\bfJ = \textbf{0}} =
\sum_{\bfm_1}...\sum_{\bfm_N}\e^{-S_0}$ and the brackets denote
thermal average with respect to $Z_0$. The functional derivation is
 performed by expanding the exponential in series and keep terms
of $\mathcal{O}(J^2)$, the only terms that survives both
derivation and $\bfJ = \textbf{0}$, to yield 

\begin{gather}
\begin{split}
\frac{\delta^2
  \e^{-S_1}}{\delta J_{\mu}(-\bfq) \delta J_{\nu}(\bfq)}\Bigg|_{\bfJ =
  \textbf{0}} &= \frac{4\pi^2e_je_lR_{jk}R_{lm}\varepsilon_{\mu \alpha
  \beta}\varepsilon_{\nu \gamma \kappa}}{q^4(q^2+m_{0}^{2})^2}\\
&\;\quad\times q_{\alpha}q_{\gamma}m_{\beta, k}(\bfq)m_{\kappa, m}(-\bfq)
\\
&\quad+\frac{P_{T, \mu \nu}}{\beta(q^2+m_{0}^{2})}.
\end{split}
\label{Eq:S_1_derivation}
\end{gather}
The product $\varepsilon_{\mu \alpha \beta}\varepsilon_{\nu \gamma
  \kappa}$ is evaluated by the determinant

\begin{equation}
\varepsilon_{\mu \alpha \beta}\varepsilon_{\nu \gamma \kappa} =
\begin{vmatrix}
\delta_{\mu \nu} & \delta_{\mu \gamma} & \delta_{\mu \kappa} \\
\delta_{\alpha \nu} & \delta_{\alpha \gamma} & \delta_{\alpha \kappa} \\
\delta_{\beta \nu} & \delta_{\beta \gamma} & \delta_{\beta \kappa} 
\end{vmatrix},
\end{equation}
to yield

\begin{equation}
\begin{split}
\langle A_{\mu}(\bfq) A_{\nu}(-\bfq)\rangle &= \frac{P_{T, \mu
    \nu}}{\beta(q^2+m_{0}^{2})} + \frac{4\pi^2e_je_lR_{jk}R_{lm}}{q^2(q^2+m_{0}^{2})^2}\\
&\quad\times \langle P_{T, \mu \nu}\bfm_{k}(\bfq)\bfm_{m}(-\bfq)\\
&\quad\qquad-m_{\nu, k}(\bfq)m_{\mu,
  m}(-\bfq)\rangle,
\end{split}
\label{Eq:General_gauge_field_propagator}
\end{equation}
when Eq. \eqref{Eq:S_1_derivation} is inserted in
Eq. \eqref{Eq:prop_definition}. We now find the gauge-field propagator by
letting $\nu \rightarrow \mu$ in Eq. \eqref{Eq:General_gauge_field_propagator} and summing 
over repeated indices, thus
\begin{equation}
\begin{split}
\langle \bfA(\bfq)  \bfA(-\bfq)\rangle =
& \frac{4\pi^2e_je_lR_{jk}R_{lm}\langle \bfm_{k}(\bfq)  \bfm_{m}(-\bfq)\rangle}{q^2(q^2+m_{0}^{2})^2}\\
&+\frac{2}{\beta(q^2+m_{0}^{2})}.
\end{split}
\label{Eq:Gauge_field_propagator}
\end{equation}
The gauge-field correlator of the 
two-component discrete model in Eq. \eqref{Eq:Villain_H} is found similarly to Appendix 
C in Ref. \onlinecite{prb05} and the result is as given in Eq. \eqref{Eq:Gauge_field_propagator} 
with $q^2 \rightarrow |\bfQ_{\bfq}|^2$ and vortex fields defined on the vertices of the 
Fourier space dual lattice.

\end{document}